\documentclass[%
 reprint,
superscriptaddress,
 amsmath,amssymb,
 aps,
pra,
floatfix,
]{revtex4-2}

\usepackage{graphicx}
\usepackage{dcolumn}
\usepackage{bm}
\usepackage{braket}
\usepackage{siunitx}
\usepackage{bbold}
\DeclareSIUnit{\belmilliwatt}{Bm}
\DeclareSIUnit{\dBm}{\deci\belmilliwatt}
\usepackage[T1]{fontenc}
\usepackage{placeins}
\usepackage[utf8]{inputenc}
\usepackage{mathtools}

\usepackage{hyperref}
\hypersetup{
    colorlinks=true,
    linkcolor=magenta,
    citecolor=magenta,
    filecolor=magenta,      
    urlcolor=magenta,
}

\begin{document}

\title{Weight-four parity checks with silicon spin qubits}

\author{Brennan~Undseth}
\thanks{These authors contributed equally to this work.}
\author{Nicola~Meggiato}
\thanks{These authors contributed equally to this work.}
\author{Yi-Hsien~Wu}
\author{Sam~R.~Katiraee-Far}
\affiliation{QuTech and Kavli Institute of Nanoscience, Delft University of Technology, Lorentzweg 1, 2628 CJ Delft, The Netherlands}
\author{Larysa~Tryputen}
\affiliation{Netherlands Organization for Applied Scientific Research (TNO), Stieltjesweg 1, 2628 CK Delft, Netherlands}
\author{Sander~L.~de~Snoo}
\author{Davide~Degli~Esposti}
\author{Giordano~Scappucci}
\author{Eli\v{s}ka~Greplov\'a}
\author{Lieven~M.~K.~Vandersypen}
\thanks{Corresponding author:
l.m.k.vandersypen@tudelft.nl}
\affiliation{QuTech and Kavli Institute of Nanoscience, Delft University of Technology, Lorentzweg 1, 2628 CJ Delft, The Netherlands}

\begin{abstract}

Recent advances in coherent spin shuttling have made sparse semiconductor spin qubit arrays an appealing solid-state platform to realize quantum processors. The dynamic and long-range connectivity enabled by shuttling is also essential for many quantum error-correction (QEC) schemes. Here, we demonstrate a silicon spin-qubit device that comprises a shuttling bus for coherently transporting qubits that can interact at four isolated locations we call bus stops. We dynamically populate the array and tune all single- and two-qubit operations using shuttling and quantum non-demolition (QND) spin measurements, without access to charge sensing in most of the device. We achieve universal control of the effective five-qubit processor and select the connectivity required to form a surface-code stabilizer plaquette that supports $X$- and $Z$-type parity checks up to weight-four. We use the parity checks to generate multi-qubit entanglement between all qubit combinations in the array and report the genuine entanglement of a five-qubit Greenberger-Horne-Zeilinger (GHZ) state, constituting the largest such state ever constructed with gate-defined semiconductor spins. This work opens immediate opportunities to pursue QEC experiments with spin qubits, and the protocols developed here lay the groundwork for the modular calibration and operation of sparse spin qubit arrays.

\end{abstract}

\maketitle

For semiconductor-based spin qubits to be a viable platform for fault-tolerant quantum computing, their physical advantages, including their small size, long coherence times, and high-fidelity operations~\cite{burkard_2023}, must be compatible with quantum error correction (QEC). Although previous demonstrations of phase-flip codes with spin qubits have made use of Toffoli-like primitives to implement conditional correction operations~\cite{Takeda_2022, van_Riggelen2022}, large-scale deployment of QEC requires the repeated extraction of stabilizer measurements to infer error syndromes in both space and time. A well-established error correction benchmark, the surface code, uses weight-four $X$- and $Z$-type parity checks to achieve quantum memories with reasonable error thresholds~\cite{Raussendorf_2007, Fowler2012}. Such operations require an ancillary qubit to interact with four data qubits, and the former must be measured without destroying the quantum state of the latter. This connectivity is challenging to realize in densely populated quantum dot arrays due to crosstalk and the co-integration of readout, but spin shuttling in sparsely-occupied arrays has long been recognized as a path forward~\cite{Taylor2005}.

The performance of semiconductor-based spin shuttling has progressed rapidly in recent years due to improvements in material uniformity and control techniques \cite{Fujita_2017, Jadot_2021, Mills_2019, Yoneda_2021, Struck_2024, Xue_2024, van_Riggelen_Doelman_2024, De_Smet_2025, ademi2025distributingentanglementdistantsemiconductor, Foster_2025, Degli_Esposti_2024}. More recently, spin shuttling has also been used as a means of material characterization \cite{Volmer_2024}, qubit characterization \cite{Wang_2024}, and implementing logical gates \cite{Noiri_2022, matsumoto2025twoqubitlogicteleportationmobile}. These advancements have maintained substantial interest in sparse spin qubit architectures for fault-tolerant quantum information processing where the physical movement of spins enables the coupling of qubits at distances far greater than kinetic exchange or capacitive coupling and allows for the dynamic reconfiguration of qubit connectivity \cite{Taylor2005, Xue_thesis,  Buonacorsi_2019, Boter_2022, Cai_2023, Kunne_2024, siegel2025snakesplanemobilelow}.

Qubit shuttling in other quantum computing platforms, such as trapped ion systems \cite{Moses_2023} and neutral atom systems \cite{Bluvstein_2023}, enables zoned architectures and permits beyond-planar connectivity that is relevant for QEC codes with higher encoding rates \cite{Breuckmann_2021}. For semiconductor-based spins in particular, the additional separation between shuttled qubits in sparse arrays offers advantages beyond higher connectivity. First, residual exchange between adjacent spins can be effectively eliminated during idling, greatly reducing the potential for entangling crosstalk \cite{Heinz_2024}. Second, a lower qubit density eases the routing and localization of control signals, which consequently reduces the calibration overhead required to mitigate capacitive crosstalk \cite{Rao_2025, Jirovec_2025}. Finally, spin qubit readout becomes more flexible as the co-integration of charge sensing can require charge reservoirs, DC current paths \cite{Kiene2025} or rf resonators \cite{Vigneau_2023} that are bulkier than the individual qubits. However, to realize these advantages of sparser architectures, coherent spin shuttling must be integrated within a multi-qubit setting while maintaining universal control and efficient measurement \cite{wu2025simultaneoushighfidelitysinglequbitgates, Xue_2022, Noiri_2022_FT, Mills_2022, Takeda_2024, Philips_2022}.

In this work, we commission a sparse five-qubit array in silicon that leverages spin shuttling to enable long-range connectivity. In contrast to previous demonstrations with semiconductor spins, we characterize the array and tune logical control using shuttling in the absence of direct charge sensing. The resulting protocol allows for modular calibration and optimization, and it illustrates a viable alternative to the challenges of experimentally realizing dense qubit arrays. After outlining the measurement framework used to initialize and readout the multi-qubit state of the device, we benchmark the quantum processor and report the successful implementation of parity checks up to weight-four as well as the creation of genuine entanglement between all qubit combinations in the array. These results highlight both the feasibility and the architectural benefits of incorporating shuttling into semiconductor quantum processors.

\begin{figure}
    \centering
    \includegraphics[width=\columnwidth]{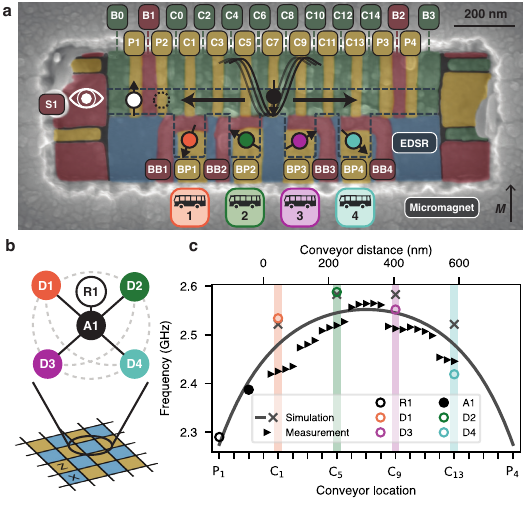}
    \caption{\textbf{Device concept and shuttling bus}. \textbf{a} False-colored scanning electron microscope image of a device nominally identical to that used in the experiments. Four metallization layers (from bottom to top: blue, red, yellow, green) are used to define a single-electron transistor charge sensor (S1), the readout zone (B0-P2), the shuttling bus (C0-C14), and four bus stops (BP1-BP4). The readout zone on the right side of the device (P3-B3) is functional but unused, and no electrons are accumulated beneath P3 or P4. A partially-magnetized cobalt micromagnet (gray) provides an inhomogeneous magnetic field profile to define and operate the qubits. No external magnetic field is present. The colored spins correspond to the qubit labels in \textbf{b}. \textbf{b} The qubit connectivity graph made possible via the shuttling bus. Solid lines indicate connections used in this work, whereby the shuttled ancilla qubit (A1) may interact with the four data qubits (D1-D4) as well as the readout ancilla (R1). This connectivity forms a surface code stabilizer plaquette in the context of a larger architecture. Dashed lines indicate connections that are physically available but unused in this work. \textbf{c} The simulated and measured Larmor frequency profile along the shuttling bus axis. Analogous values for a spin localized under the bus stop plungers (BP1-BP4) are plotted at the adjacent shuttling bus gates (C1-C13) and highlighted by the respective colors depicted in \textbf{a}.}
    \label{fig1}
\end{figure}

The sparse array, as shown in Fig.~\ref{fig1}a, is electrostatically defined in the isotopically purified quantum well of a $^{28}$Si/SiGe heterostructure (see Sec.~\ref{supp:device}) and consists of three zones: a readout zone, a spin shuttling bus, and an adjacent row of quantum dots we refer to as bus stops. At constant DC voltages, only the quantum dots formed under gates P1 and P2 are populated with 3 and 1 electrons, respectively, to form a qubit readout pair using parity-mode Pauli spin blockade (PSB) for spin-to-charge conversion  \cite{Seedhouse_2021}. The adjacent sensor S1 is a single-electron transistor (SET) connected to an off-chip tank circuit such that rf-reflectometry may be used to probe the charge state of the readout pair (see Sec.~\ref{supp:PSBreadout}). The spin qubit R1, kept constantly below gate P1, is only used as a readout ancilla in this work, although it is fully controllable. The qubit A1, initialized below gate P2, functions as a mobile ancilla with which to characterize the sparse array and connect the qubits within it.

The overlapping gates C0-C14 in the top row of the array define the shuttling bus in the channel between the screening gates. The bottom row defines the bus stops with four individual plunger gates BP1-BP4 to accumulate four spin qubits, respectively D1-D4, and four adjacent barrier gates BB1-BB4 to control the tunnel couplings to the bus. The resulting connectivity graph, shown in Fig.~\ref{fig1}b, corresponds to a weight-four stabilizer plaquette where each data qubit may directly interact with the ancilla qubit by coherently shuttling the latter to a position adjacent to the data qubit bus stop. The data qubits could also be shuttled and interact directly with one another for effective all-to-all connectivity~\cite{Xue_thesis}, but we do not make use of these couplings in this work. It has been shown that extending such a bilinear architecture is conceptually sufficient to support the surface code \cite{Siegel2024}, while the addition of more shuttling channels increases the architectural flexibility further \cite{Cai_2023}.

The gate stack consists of four metallization layers and an on-chip micromagnet. The bottom screening gate is used as a microwave antenna for driving resonant single-qubit gates and conditional rotations (CROT) via electric-dipole spin resonance (EDSR)~\cite{burkard_2023}. The gate layer order is selected (refer to Fig.~\ref{fig1}a) to maximize the tunnel coupling tunability for two-qubit interactions between the ancilla and data qubits as well as the interaction between the PSB pair, as gates in higher layers typically exhibit lower lever arms to the buried quantum well \cite{park2025highlytunabletwoqubitinteractions}. Each labeled finger gate is controlled individually, although the experiments presented here are conceptually compatible with shared control of the shuttling bus \cite{Struck_2024, ademi2025distributingentanglementdistantsemiconductor}.

The on-chip cobalt micromagnet is used to engineer the spin Hamiltonian of the array (see Sec.~\ref{supp:magnetostatics}). After magnetizing the micromagnet in an external field of \SI{1}{\tesla}, the external field is turned off so that only the micromagnet remanence field remains. Fig.~\ref{fig1}c shows the simulated parabolic trend in Larmor frequencies of spins located along the bus axis along with the estimated Larmor frequencies of spins localized in the adjacent bus stops. These frequencies are experimentally probed by shuttling the ancilla spin prepared in a superposition with a resonant burst and measuring the relative change in Larmor precession. We use a four-phase traveling-wave potential to propagate spins through the bus~\cite{Seidler_2022,De_Smet_2025,ademi2025distributingentanglementdistantsemiconductor},
at a speed of \SI{1.8}{\meter/\second} (see Sec.~\ref{supp:conveyor}). The sudden steps in the measured trend indicate confinement potential roughness \cite{Langrock2023}. The shuttling operation is adiabatic, and the effective unitary acting on the spin is a phase pickup that can be calibrated explicitly or negated using a spin echo.

\begin{figure*}
    \centering
    \includegraphics[width=\textwidth]{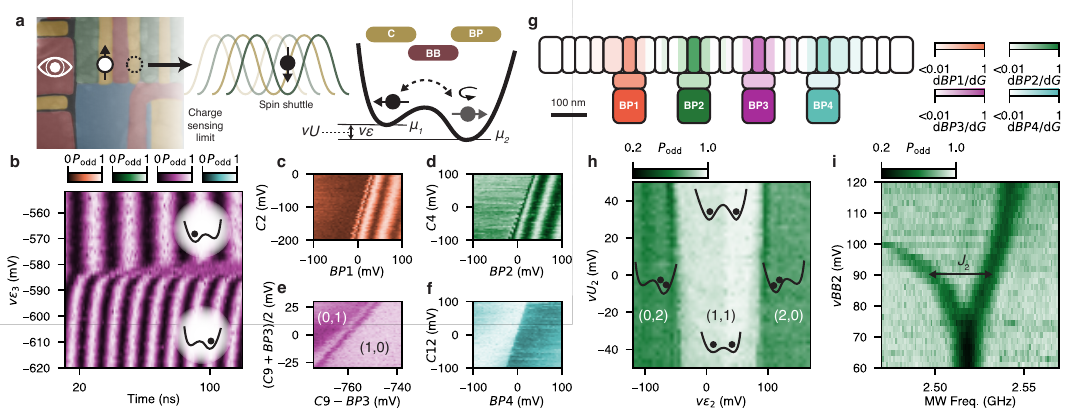}
    \caption{\textbf{Remote tuning protocols}. \textbf{a} Illustration of the remote tuning concept. The sensor has a limited charge sensing range and is therefore only used for qubit measurement of the ancilla spin. To probe control of the bus stops, the ancilla is shuttled beyond charge sensing range, and spin physics is used to calibrate the virtual detuning $v\epsilon \propto \mu_1-\mu_2$ and average chemical potential $vU\propto \left(\mu_1+\mu_2\right)/2$ for all four effective double-dot systems formed at the bus stops (see Sec.~\ref{supp:1etuning}). \textbf{b} An example of a Ramsey tunneling experiment at bus stop 3. Coherent tunneling of the ancilla spin from the shuttling bus to the bus stop is recognized by the abrupt change in Larmor precession frequency. Colorbars are shared with \textbf{c}-\textbf{f}. \textbf{c}-\textbf{f} Examples of charge transitions identified via Ramsey tunneling experiments for bus stops 1-4 respectively with wait times on the order of tens of nanoseconds. The interdot charge transition between the shuttling bus and bus stop 3 which may be used to infer control of $v\epsilon$ and $vU$ is shown in \textbf{e}, while the other scans are used to extract other virtual gate matrix elements. \textbf{g} Crosstalk heatmap with respect to the interdot charge transition between the bus and each bus stop extracted from scans analogous to those in \textbf{c}-\textbf{f} (see Fig.~\ref{fig:sparse_crosstalk}). Detectable crosstalk to multiple bus stops is indicated by split coloring. \textbf{h} An effective two-electron charge stability diagram of the isolated double dot system formed at bus stop 2 identified from the returned polarization of a shuttled spin. \textbf{i} Detection of the exchange interaction strength $J_2$ between a shuttled spin and a mixed spin state in bus stop 2 obtained by applying microwave bursts at the charge symmetry point as determined from \textbf{h}. See Fig.~\ref{fig:bus_stop_calibrations} for analogous calibrations of all four bus stops.}
    \label{fig2}
\end{figure*}

The conventional method of tuning spin qubit arrays begins with using a charge sensor to tune quantum dots into known charge states, followed by using spin physics to calibrate and refine qubit interactions. In this device, direct charge sensing with S1 is possible only up to bus stop 1, so we pioneer a remote tuning procedure that allows us to populate and virtualize control of the distant bus stops as well as tune single- and two-qubit operations without local charge sensing. Fig.~\ref{fig2}a provides a conceptual illustration where a spin is placed in a superposition and shuttled to a bus stop out of charge sensing range. After pulsing the effective double-dot system to a particular balance of chemical potentials, the phase of the resulting superposition will depend on whether coherent tunneling between dots occurred. This phase is measured by shuttling the spin back to the readout zone. Fig.~\ref{fig2}b depicts an instance of this tunneling, similar to measurements performed in Si-MOS and Ge/SiGe spin qubits while in range of charge sensing \cite{Yoneda_2021, van_Riggelen_Doelman_2024}.

By fixing the time of free evolution, single-electron charge stability diagrams are reconstructed from the shuttled spin phase as in Fig.~\ref{fig2}c-f for all four bus stops. Based on the slope of the interdot transition, the relative capacitive coupling between all gates and the bus stop quantum dot can be inferred and compensated through gate virtualization (see Sec.~\ref{supp:1etuning}). Fig.~\ref{fig2}g illustrates a heatmap of all such electrostatic crosstalk and highlights one of the benefits of increased dot separation. While a few intermediate gates in the shuttling bus have a weak effect on two bus stops, the charge transition to each bus stop can be treated as effectively independent, allowing modular charge tuning of the array. Virtualization of the gates in the shuttling bus with respect to the bus stop plungers prevents the occurrence of unintended electron transitions to the bus stops during shuttling. However, the bus stop plunger gates are not virtualized with respect to the gates of the shuttling bus. We therefore expect the electrostatic potential in the shuttling channel to change when gate voltages are pulsed to accumulate data qubits, and we use two different conveyor definitions for charge loading and multi-qubit operation to account for this (see Sec.~\ref{supp:conveyor}).

After a spin has shuttled through the bus and tunneled into a bus stop, we can load a new spin into the array on-demand from the sensor S1, which doubles as an electron reservoir, in approximately \SI{10}{\micro\second} (see Sec.~\ref{supp:reload}). To calibrate two-qubit logic between the ancilla qubit in the bus and a data qubit in a bus stop, the correct electrostatic balance in the effective double-dot system must be ensured. To do this, we initialize the ancilla to $\ket{\downarrow}$ after loading the bus stop with a mixed spin state. The ancilla spin is shuttled to be adjacent to the bus stop for \SI{10}{\micro\second}, and the polarization of the returned spin serves as a probe of whether tunneling to or from the bus stop and subsequent spin mixing occurred. The resulting spin signal can be used to infer two-electron charge stability diagrams as shown in Fig.~\ref{fig2}h. We note that such a method could form the basis for more rigorous spectroscopy of excited valley-orbital states far from local charge sensing as well \cite{Hanson_2007}.

After identifying the approximate symmetry point ($v\epsilon=0$) of the two-electron charge state in each bus stop (see Fig.~\ref{fig:bus_stop_calibrations}i-l), the exchange splitting is probed by modulating the bus stop barrier gates to establish a suitable range for two-qubit interactions as shown in Fig.~\ref{fig2}i. Exchange tunability on the order of \SI{10}{\mega\hertz} is suitable for resonant controlled rotations (CROT) and adiabatic controlled-phase (CZ) operations~\cite{burkard_2023}, both of which provide a universal entangling gate. Verifying the tunability of all four unique exchange interactions after sequentially loading all four bus stops confirms that the loaded spins are robust to the interleaved shuttling operations and suitable for calibrating quantum logic (see Fig.~\ref{fig:bus_stop_calibrations}m-p).

\begin{figure*}
    \centering
    \includegraphics[width=\textwidth]{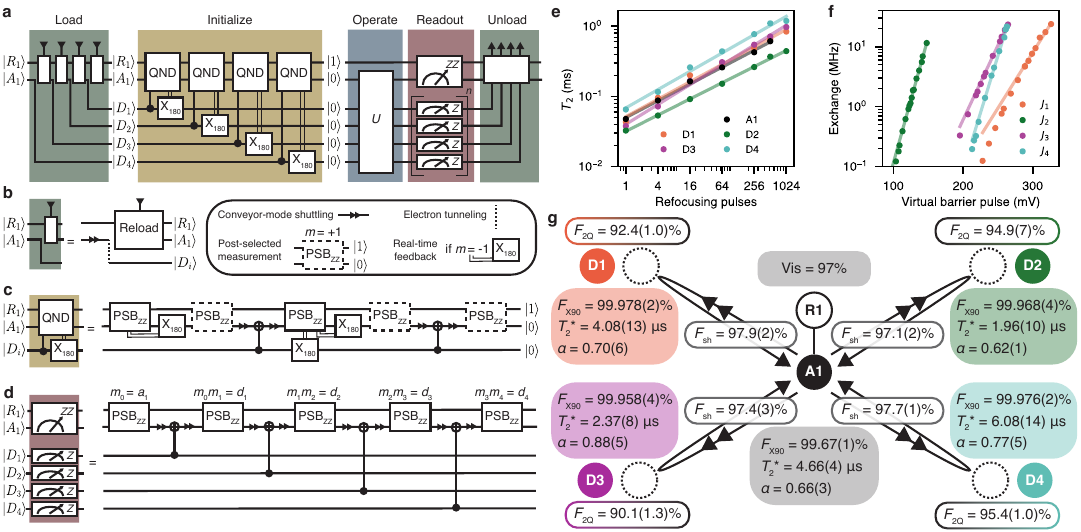}
    \caption{\textbf{QND measurement framework and QPU characterization}. \textbf{a} Schematic showing the overall protocol for loading, initializing, measuring, and unloading the sparse array when used as a quantum processor. \textbf{b} Building block of the loading protocol. The ancilla qubit is shuttled along the bus, tunnels into the selected bus stop, and is relabeled as a data qubit. A new ancilla is then reloaded from S1 as described in Sec.~\ref{supp:reload}. \textbf{c} Building block of the initialization protocol. Two rounds of QND measurements are performed on the data qubit, with real-time feedback used to correct the measured state of both the ancilla and data qubits. Post-selection is used to further boost the initialization fidelity by filtering certain error modes, such as infidelity in the CROT. A variable number of data qubits can be initialized in any order using this fixed building block at the expense of some redundancy, but this does not cause a meaningful performance bottleneck. \textbf{d} Building block of the QND readout protocol. Extraction of the five-qubit computational-basis measurement is elaborated in Sec.~\ref{supp:QNDmeasurement}. \textbf{e} CPMG coherence time measured for the ancilla qubit and four data qubits. Fits are according to an assumed $1/f^\alpha$ noise power spectral density (see Sec.~\ref{supp:1Qcharacterization}). \textbf{f} Exchange tunability of all four interactions between data qubits and the ancilla. Fits assume an exponential relation between barrier voltage and exchange. \textbf{g} Overview of all benchmarked operations and qubit properties in the five-qubit processor (see Sec.~\ref{supp:1Qcharacterization} and \ref{supp:2Qcharacterization}).}
    \label{fig3}
\end{figure*}

To operate the sparse spin qubit array as a five-qubit quantum processor, the protocol shown in Fig.~\ref{fig3}a is used. First, all four data qubits are loaded into the bus stops in mixed states via shuttling the ancilla qubit and reloading as depicted in subcircuit Fig.~\ref{fig3}b. After reloading the ancilla qubit for the final time, a QND protocol, expanded in Fig.~\ref{fig3}c, is used to initialize both the ancilla and data qubits by measurement and using real-time feedback \cite{Philips_2022, Huang_2024}. Direct initialization of the data qubits at the readout zone prior to shuttling is possible, but the QND protocol generally permits higher measurement fidelities due to its repeatability. At the end of the initialization sequence, all spins are initialized in the ground state $\ket{\downarrow} \equiv \ket{0}$ with the exception of R1, which is initialized to $\ket{\uparrow} \equiv \ket{1}$. After universally controlling the spins with a combination of coherent shuttling, single-qubit EDSR, and two-qubit exchange, they are measured using a generalized multi-qubit QND measurement, shown in Fig.~\ref{fig3}d. The measurement yields up to five bits of information corresponding to the computational-basis measurement of the ancilla qubit and the four data qubits, and multiple rounds of QND measurement may be used to improve the readout fidelity of the data qubit states (see Sec.~\ref{supp:QNDmeasurement}). After measurement, the pulsed gate voltages are returned to zero and the bus stop spins are unloaded before a next experimental cycle is initiated (see Sec.~\ref{supp:reload}). The dominant timescale is the \SI{10}{\micro\second} integration time associated with each PSB measurement, and the minimal duration of an experimental cycle is about \SI{300}{\micro\second}.

We use the QND measurement framework to calibrate and characterize all relevant operations for multi-qubit control using established methods (see Sec.~\ref{supp:universalcontrol} and \ref{supp:gatecalibration}). With dynamical decoupling, all qubits maintain coherence for hundreds of microseconds and exhibit behavior consistent with a $1/f^\alpha$ noise power spectral density as seen in Fig.~\ref{fig3}e. We note that the data qubit coherence far exceeds the \SI{10}{\micro\second} readout time and permits coherence-preserving mid-circuit measurements in this device \cite{jones2025midcircuitlogicexecutedqubit}. Fig.~\ref{fig3}f illustrates the dynamic range of exchange tunability that is used to implement entangling two-qubit operations. Residual exchange between spins loaded in the bus stops is directly verified to be at least below \SI{10}{\kilo\hertz}, though it is likely much smaller, and therefore entangling crosstalk within the sparse array is negligible (see Fig.~\ref{fig:residual_exchange}). We note that the residual exchange between data qubits and the ancilla is not relevant as the ancilla never idles adjacent to any bus stop.

Fig.~\ref{fig3}g illustrates all benchmarks when the device is fully operational. Although the single-qubit fidelities $F_{X90}$ are respectable when compared with the state-of-the-art \cite{wu2025simultaneoushighfidelitysinglequbitgates}, we find that the two-qubit gate fidelities $F_{2Q}$ are limited by incoherent noise in the exchange interaction strength (see Fig.~\ref{fig:exchangeQ}). Previous demonstrations in two-qubit devices show that the technology is capable of two-qubit gate fidelities well over 99\% \cite{Tanttu_2024, Xue_2022, Noiri_2022_FT, Mills_2022, Steinacker_2025}, and it will require ongoing work to routinely incorporate higher-fidelity two-qubit gates in a multi-qubit setting.

We characterize the \SI{1.2}{\micro\meter} round-trip shuttling path to have a single-qubit fidelity of 97.7(1)\% with respect to an identity operation when the array is operated as a five-qubit processor. The largest error contribution is associated with the charge transition from below P2 to C0. We estimate the remaining infidelity to be consistent with dephasing and discuss the possible role of valley excitations further in Sec.~\ref{supp:1Qcharacterization}. Although higher shuttling fidelity has been reported over a larger cumulative distance \cite{De_Smet_2025}, our results demonstrate the longest coherent spin shuttle over a real distance in silicon. While further refinement of the traveling-wave potential would likely enhance both the speed and fidelity of shuttling by reducing dephasing, the total error budget for operating all five qubits is dominated by the two-qubit interactions (see Sec.~\ref{supp:errorbudget}).

\begin{figure*}
    \centering
    \includegraphics[width=\textwidth]{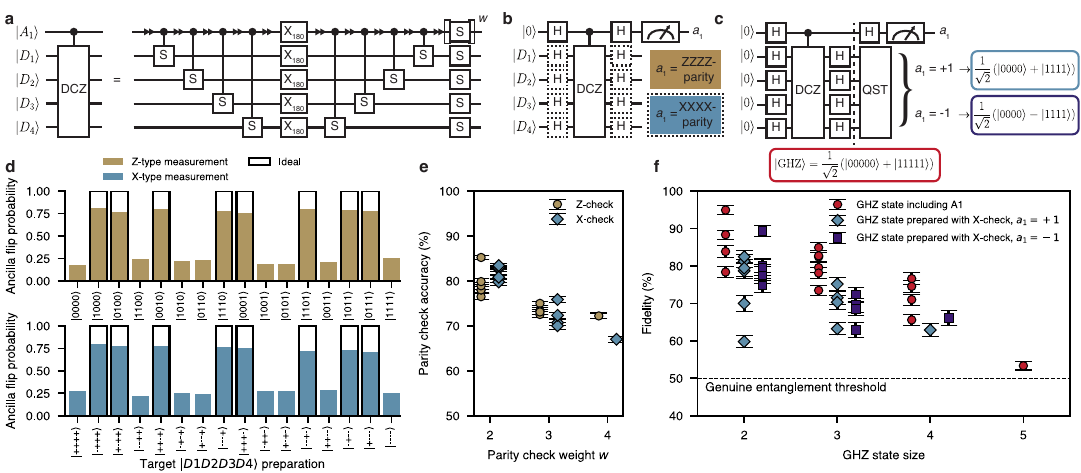}
    \caption{\textbf{Benchmarking parity checks and multi-qubit entanglement}. \textbf{a} Quantum circuit depicting the multi-qubit decoupled CZ used for coherent two-qubit operations. The circuit generalizes to any number $w$ of data qubits by omitting or adding quantum wires of different data qubits. Single-qubit gates are parallelized pairwise when relevant, and the $2w$ $S$ gates are applied virtually. \textbf{b} Quantum circuit depicting a weight-four $Z$-type parity check utilizing the multi-qubit decoupled CZ from \textbf{a}. The circuit generalizes to any number of data qubits by changing the weight $w$ of the decoupled CZ. An $X$-type check is performed with the inclusion of the dashed Hadamard gates. \textbf{c} Quantum circuit depicting $w$-qubit GHZ state generation using an $X$-type parity check. The dashed line indicates when the ancilla forms a $w+1$-qubit GHZ state with the data qubits. The outcome of the ancilla measurement determines which of two GHZ states are realized in the data qubits. One of the two outcomes can be used as a logical state of a four-qubit quantum error correction code. \textbf{d} Outcomes of the weight-four $Z$-type ($X$-type) parity check obtained after preparing all $Z$-basis ($X$-basis) eigenstates of the data qubits. The black wireframes indicate the ideal result for each input state, and we estimate that data qubit initialization errors contribute a discrepancy of 2\% in the observed ancilla outcome. All other parity checks are shown in Fig.~\ref{fig:paritychecks}. \textbf{e} The parity check accuracy for all $Z$- and $X$-type parity checks of weights two through four computed as the average probability that the ancilla measurement is correct and the data qubits are measured in the intended prepared state. Data qubit initialization and readout errors are removed by rescaling, and error bars signify $\pm1\sigma$ as determined by Monte Carlo bootstrapping (see Sec.~\ref{supp:paritychecks}). \textbf{f} GHZ state fidelities for all combinations of two or more qubits in the five qubit processor (some overlap closely). State fidelities including the ancilla are calculated with respect to the GHZ state $\frac{1}{\sqrt{2}}(\ket{0}^{\otimes (w+1)}+\ket{1}^{\otimes (w+1)})$. For GHZ states initialized via a weight-$w$ parity check, the state fidelities are calculated with respect to $\frac{1}{\sqrt{2}}(\ket{0}^{\otimes w}\pm\ket{1}^{\otimes w})$. Error bars signify $\pm 1\sigma$ as determined by Monte Carlo bootstrapping (see Sec.~\ref{supp:QST}). All reconstructed density matrices are shown in Fig.~\ref{fig:rhos_Q2}, Fig.~\ref{fig:rhos_Q3}, Fig.~\ref{fig:rhos_Q4} and Fig.~\ref{fig:rhos_Q5} along with the estimated fidelities in the absence of initialization and readout error removal.}
    \label{fig4}
\end{figure*}

To demonstrate both phase-coherent two-qubit interactions and the utility of the shuttling-enabled connectivity across the full system, we demonstrate up to weight-four parity checks of the data qubits and five-qubit entanglement. A central consideration for operating multi-qubit registers with shuttling is that moved spins will accrue phase in the rotating frame in which they are being resonantly controlled. In this device, the ancilla qubit will pick up a phase as it travels in the inhomogeneous magnetic field. The data qubits, although stationary, also pick up a phase due to the stray electric fields from the shuttling pulses coupling to the spins via the magnetic field gradient. These phases depend on the shuttling path and the number of two-qubit interactions carried out on the way, and therefore may lead to a large overhead in the number of calibrated parameters.

To avoid this, we perform two-qubit logic using a decoupled CZ gate shown in Fig.~\ref{fig4}a. In addition to decoupling low-frequency noise, the $X_{180}$ gates also refocus the single-qubit phase pickups acquired during both shuttling and two-qubit gates provided the operation is symmetric. The pairs of native conditional-$S$ gates are implemented by adiabatically modulating the exchange interactions, which commute with the refocusing pulses, and enable a maximally entangling operation between the ancilla and any combination of the data qubits depending on which two-qubit interactions are enabled. The single-qubit operation on the ancilla has the highest quality when the spin is localized below gate P2, and therefore the implemented decoupled CZ operation employs two rounds of shuttling. However, this is not a fundamental requirement of the approach.

The parity of any multi-qubit Pauli operator acting on any subset of the data qubits can be extracted with the appropriate change in basis of the data qubits as shown in Fig.~\ref{fig4}b. Fig.~\ref{fig4}c shows how a weight-$w$ $X$-type parity check acting on an input state $\ket{0}^{\otimes w}$ is used to probabilistically generate maximally entangled $w$-qubit states that differ by a single-qubit Pauli operation. In the case of a weight-four parity check, one of the post-selected states is the logical state of a $\left[\!\left[ 4, 1, 2 \right]\!\right]$ surface code while the other state is outside of the code space. We can therefore benchmark the quality of the parity checks in two ways. First, we ensure that all eigenstates of $Z$- and $X$-type parity observables are detected and preserved properly. Second, we generate multi-qubit entanglement generated via the parity checks as a more stringent test of the parity check performance.

Fig.~\ref{fig4}d shows the measured ancilla state when weight-four $Z$- and $X$-type parity checks act on all of the respective input eigenstates. Imperfections between the ideal ancilla outcome and the measured parity are a result of imperfect state preparation of ancilla and data qubits, finite ancilla measurement fidelity, finite ancilla shuttling fidelity, and finite single- and two-qubit gate fidelity. Additionally, we measure the data qubit state after the parity check and calculate the parity check accuracy as the probability that both the ancilla outcome is correct and the intended prepared state is observed. Fig.~\ref{fig4}e tabulates this parity check accuracy for all data qubit combinations. The weight-four parity check accuracy of 72.2(6)\% (67.0(7)\%) therefore establishes the quality with which the parity of the $ZZZZ$ ($XXXX$) eigenbasis can be detected and conserved. The primary difference between $X$- and $Z$-type checks is the larger number of single-qubit rotations required in the former case. In Sec.~\ref{supp:paritychecks}, we show an example of performing 35 repeated weight-two parity checks with a retention rate above 93\%.

When benchmarking the check performance on eigenstates, the data qubits are ideally $\ket{0}$ or $\ket{1}$ throughout the decoupled CZ operation and, due to the long $T_1$ times of spin qubits, robust to decoherence. The generation of entanglement is therefore a stronger test of the parity check performance, as the quantum state of the system will be maximally sensitive to operational errors, including the error rates associated with spin shuttling and crosstalk. Fig.~\ref{fig4}f tabulates the GHZ state fidelities of all combinations of entangled states possible in the five-qubit processor. After accounting for measurement errors, the observed fidelities are comparable to the best multi-qubit entangled states generated with gate-defined semiconductor spins \cite{Philips_2022, defuentes2025runningsixqubitquantumcircuit, Takeda_2021_GHZ}, and we observe genuine five-qubit entanglement ($F_\mathrm{GHZ}>50\%$) for the first time in the platform.

By postselecting the state of the ancilla measurement, we probabilistically initialize the data qubits into the logical state of a \(\left[\!\left[ 4, 1, 2 \right]\!\right]\) surface code with a state fidelity of 62.9(1.8)\%. This central result highlights that our implementation of parity checks is suitable for more sophisticated QEC experiments \cite{Andersen_2020}. We note that this procedure requires five-qubit entanglement, and we demonstrate higher four-qubit GHZ state fidelities up to 76.6(1.6)\% with qubit combinations that include the ancilla. In Sec.~\ref{supp:errorbudget}, we formulate a budget of all characterized error sources and conclude that the majority of the error originates from infidelity in the two-qubit exchange interactions. Dephasing during shuttling and idling constitutes about a quarter of the total error, and increasing the speed of shuttling is the most direct means to lower this contribution.

The incorporation of spin shuttling into a multi-qubit setting has implications both for how semiconductor spin-based quantum processors are operated and the level of performance they can achieve. We have shown that shuttling can be used to tune universal multi-qubit control without the need for local charge sensing throughout an array. The presented tuning protocols are only limited by the distance over which the shuttling operation maintains spin coherence. With high-fidelity shuttling already demonstrated over a \SI{10}{\micro\meter} length scale \cite{De_Smet_2025}, we expect that the approach presented here is sufficient to tune on the order of 100 qubits using a single charge sensor and ohmic contact in prototype industrial layouts \cite{li2025trilinearquantumdotarchitecture}. Due to the lower degree of electrostatic crosstalk in sparse arrays, this calibration would be modular and scale linearly with the number of qubits in the worst case. The addition of more shuttling channels further increases architectural flexibility and will be essential to avoid a measurement bottleneck.

Despite the finite fidelity of shuttling, we have shown that generating and preserving multi-qubit entanglement is not only versatile but was achieved over more spins than in past experiments. The connectivity enabled by shuttling allows for QND parity check measurements to be performed on flexible combinations of nonlocal data qubits. The device studied here supports the necessary operations required to implement a minimal \(\left[\!\left[ 4, 1, 2 \right]\!\right]\) code with pipelined syndrome extraction of the three necessary checks \cite{Versluis2017}. The successful implementation of repeated parity checks will require further enhancement of the parity check quality. In particular, the low two-qubit gate quality observed in this device is limited by incoherent noise, but the methods used here are fully compatible with achieving error rates below 1\%. 

Utilizing the operational right side of the device as a second ancilla and readout zone, complete single-shot syndrome extraction of a \(\left[\!\left[ 4, 2, 2 \right]\!\right]\) toric code could be explored \cite{Linke_2017}. Zero-distance state-preservation protocols would also be possible \cite{Kelly_2015, Zhang_2025}. 
The non-planar connectivity possible in larger sparse devices would permit more exotic encodings as well. The success of future implementations of quantum memories with gate-defined semiconductor spins depends on systematically combining the high fidelities and long coherence times that are now routinely demonstrated in smaller and industrially-fabricated devices with the versatile and extensible control possible in sparse architectures.

\textbf{Acknowledgements} We acknowledge useful discussions with Lars Schreiber, Yuta Matsumoto, Maxim De Smet, Sean van der Meer, Sasha Ivlev, Irene Fern\'{a}ndez de Fuentes, Eline Raymenants, Maximilian Rimbach-Russ, and Xiao Xue. We also thank Yuta Matsumoto and Maximilian Rimbach-Russ for critical feedback on the manuscript. We are grateful to Olaf Benningshof and Erik van der Wiel for providing cryogenic support and for designing the cold finger. We are also grateful to Yannick van der Linden for designing and testing the PCB and Qblox for support in using the control electronics. We also thank the Groove Quantum team for many instances of collaborative setup debugging as well as the remainder of the spin qubit groups at QuTech for helpful input and discussions. This publication is part of the `Quantum Inspire – the Dutch Quantum Computer in the Cloud' project (with project number [NWA.1292.19.194]) of the NWA research program `Research on Routes by Consortia (ORC)', which is funded by the Netherlands Organization for Scientific Research (NWO). This research was supported by the Horizon Europe program of the European Union under grant agreement no. 101174557 (QLSI2). Research was sponsored by the Army Research Office and was accomplished under Award Number: W911NF-23-1-0110. The views and conclusions contained in this document are those of the authors and should not be interpreted as representing the official policies, either expressed or implied, of the Army Research Office or the U.S. Government. The U.S. Government is authorized to reproduce and distribute reprints for Government purposes notwithstanding any copyright notation herein. This research was sponsored in part by The Netherlands Ministry of Defense under Awards No. QuBits R23/009. The views, conclusions, and recommendations contained in this document are those of the authors and are not necessarily endorsed, nor should they be interpreted as representing the official policies, either expressed or implied, of The Netherlands Ministry of Defense. The Netherlands Ministry of Defense is authorized to reproduce and distribute reprints for Government purposes notwithstanding any copyright notation herein..

\textbf{Author Contributions} B.U., N.M, Y-H.W. and S.K-F. performed the experiments. B.U. and N.M. built the experimental setup. B.U. designed the device and L.T. fabricated the device with substantial input from D.D.E.. S.L.S. developed the software used in the experiments. G.S. provided the $^{28}$Si/SiGe heterostructure. B.U. and L.M.K.V. conceived of the project. E.G. and L.M.K.V. supervised the project. B.U., N.M., Y-H.W. and L.M.K.V wrote the manuscript with input from all authors.

\textbf{Competing interests} The authors declare no competing interests.

\textbf{Data availability} Data supporting this work are available at Zenodo, https://doi.org/10.5281/zenodo.17368102.

\section*{supplementary information}
\appendix

\section{Device fabrication}
\label{supp:device}
The device is fabricated on an isotopically purified $^{28}$Si/SiGe heterostructure containing a \SI{7}{\nano\meter} thick strained quantum well \cite{Degli_Esposti_2024}, \SI{30}{\nano\meter} SiGe buffer passivated with an amorphous silicon cap, and a \SI{10}{\nano\meter} atomic layer deposition of insulating Al$_2$O$_3$. The heterostructure is nominally equal to that in previous studies where large average valley splitting in excess of \SI{200}{\micro\eV} \cite{Degli_Esposti_2024} and high fidelity shuttling \cite{De_Smet_2025} have been observed. The four proceeding Ti:Pd layers have thicknesses of 3:17, 3:27, 3:27, and 3:27 \SI{}{\nano\meter} respectively and are each followed by \SI{5}{\nano\meter} atomic layer deposition of Al$_2$O$_3$. Finally, a 3:150 \SI{}{\nano\meter} Ti:Co micromagnet is evaporated.

\section{Experimental setup}
\label{supp:setup}

The experiments are performed in an Oxford Instruments ProteoxMX dilution refrigerator where the mixing chamber temperature is held at \SI{200}{\milli\kelvin} to mitigate heating effects \cite{Undseth_2023}. A driven superconducting vector magnet is used to magnetize the on-chip micromagnet in a field of \SI{1}{\tesla} and is otherwise unpowered. The device is glued with GE varnish to a copper plate in direct thermal contact with the mixing chamber and is wirebonded to an in-house PCB. We apply DC bias voltages supplied by battery-powered home-built voltage source modules (D5a) and generate baseband control and readout pulses with a Qblox Cluster equipped with 8 4-channel QCM AWG modules and a QRM module for rf-reflectometry readout. DC channels are filtered with a combination of PI and RC filters with a nominal cutoff frequency of about \SI{20}{\hertz}. AC channels are filtered by ferrite chokes, pass through UT85 stainless steel semi-rigid coaxial cables, and are attenuated by \SI{20}{\deci\bel} at \SI{4}{\kelvin} to balance thermalization and having a large dynamic range for baseband pulsing. The ±\SI{2.5}{\volt} output range of the AWG channels corresponds to a ±\SI{500}{\milli\volt} available range at the device. The DC and AC signals are combined through bias tees with a time constant of \SI{100}{\milli\second}. Two channels of a QCM module are used for IQ modulation of a Rohde \& Schwarz SGS100A vector source. We set the local oscillator of the vector source to \SI{2.27}{\giga\hertz} to ensure that all target qubit frequencies fall on the same sideband within the \SI{400}{\mega\hertz} bandwidth of the QCM. The output power of the vector source is set to \SI{6}{\deci\belmilliwatt}.

The rf tone for readout is generated by the QRM, attenuated by \SI{40}{\deci\bel} at room temperature and \SI{20}{\deci\bel} at \SI{4}{\kelvin}, and the input and output signals pass through a MiniCircuits ZEDC-15-2B directional coupler at the mixing chamber stage. The input power is set to optimize the signal-to-noise ratio (SNR) of charge sensing. The output signal is carried through a superconduting NbTi UT85 cable to a Cosmic Microwave Technologies CMT-BA1 cryogenic amplifier mounted at the \SI{4}{\kelvin} stage which provides approximately \SI{30}{\deci\bel} of amplification. The signal is further amplified at room-temperature with a home-built amplifier (standalone M2j) which provides an additional \SI{45}{\deci\bel} of amplification before being filtered, digitized, and demodulated by the QRM. The demodulated signal is digitally rotated to the in-phase quadrature to make use of real-time feedback.

\section{Ancilla readout}
\label{supp:PSBreadout}

Charge sensing is achieved via an rf-SET measurement. The 2DEG leads of each SET are accumulated according to the split-gate method, whereby the rf signal is capacitively coupled to the 2DEG via an accumulation gate to miminimize series resistance and remove leakage pathways \cite{Liu_2021}. An LC tank circuit is formed with a NbTiN meandering superconducting nanowire with a nominal kinetic inductance on the order of a few \SI{}{\micro\henry}, the series capacitance between the accumulation gate and the 2DEG, and the parasitic capacitance between the bond wires and the PCB ground plane, giving rise to a resonance frequency of \SI{114.3}{\mega\hertz}.

Spin to charge conversion is achieved via parity-mode PSB. The Zeeman energy difference of about \SI{100}{\mega\hertz} between the readout ancilla R1 and ancilla A1 lifts blockade of the $\ket{T_0}$ spin triplet, resulting in an effective $Z\otimes Z$ observable for the qubit pair \cite{Seedhouse_2021}. The (4,0)-(3,1) interdot charge transition is used such that R1 occupies the dot where the lowest valley-orbit shell is filled, increasing the energy required to lift blockade. A charge stability diagram of this region can be seen in Fig.~\ref{fig:reload}b.

PSB is first tuned in isolation-mode, where the tunnel barrier to the reservoir (S1) is completely closed, as this makes the visual characteristic easier to recognize during manual tuning with video-mode charge state measurements while sweeping the respective plunger gates of the PSB pair. After initial identification of the qubit resonances, PSB is retuned with a finite reservoir coupling to allow for efficient qubit reloading.

The tunnel coupling between the quantum dots of the readout pair is pulsed to be on the order of several \SI{}{\giga\hertz} such that the singlet state $\ket{S(4,0)}$ evolves adiabatically to the $\ket{\uparrow\downarrow}$ spin state of the (3,1) configuration with a \SI{50}{\nano\second} linear voltage ramp. This allows for the initialization of an unblockaded $\ket{\uparrow\downarrow}$ state via post-selection. Real-time feedback serves to enhance the fraction of post-selected measurement shots by flipping the state of A1 if blockade is detected during initialization. Although we conceptually identify the ancilla qubit as the single spin A1 that is shuttled through the device, it is technically correct to attribute the qubit to the parity of the $\ket{R1}\otimes\ket{A1}$ spin state. The initialization and readout protocols of Fig.~\ref{fig3} only require PSB to be parity-preserving to function correctly.

After achieving reasonable readout visibility with manual tuning, the visibility of Rabi oscillations is used as a cost function for a CMA-ES optimizer to maximize the quality of initialization and readout \cite{katiraeefar2025unifiedevolutionaryoptimizationhighfidelity}. The optimizer takes as parameters the voltage pulse amplitudes on all gates immediately surrounding the double quantum dot system hosting the readout pair, as well as the durations of piecewise-linear voltage ramps defining the readout sequence. We routinely achieve ancilla qubit visibilities of about 97\% with such an optimization, corresponding to initialization and readout error rates below 1\%, while using an integration time of \SI{10}{\micro\second}.

\section{Conveyor-mode shuttling}
\label{supp:conveyor}

To generate the traveling-wave potential for the shuttling bus, we make use of a four-phase conveyor-mode pulse template as in \cite{Langrock2023, Struck_2024, Seidler_2022, De_Smet_2025,ademi2025distributingentanglementdistantsemiconductor}. During conveyor operation, every participating virtual gate $\mathrm{vG}
\in\{\mathrm{vC}0,\dots\,\mathrm{vC}12,\mathrm{vU}_1,\mathrm{v\epsilon}_1,\dots,\mathrm{vU}_4,\mathrm{v\epsilon}_4\}$ is assigned a voltage $vG$ such that:

\begin{equation}
    vG(t) = vG^\mathrm{offset}+vG^\mathrm{amp}\sin(2\pi f_\mathrm{conv} t-\phi_G),
\end{equation}

\noindent where $\phi_G = m\pi/2$ for integer $m = G~\mathrm{mod}~4$ such that every fourth virtual gate is in-phase. $\mathrm{vU}_{1-4}$ and $\mathrm{v\epsilon}_{1-4}$ are used in place of $\mathrm{vC1}$, $\mathrm{vC5}$, $\mathrm{vC9}$, and $\mathrm{vC13}$ to avoid unwanted charge transitions at the bus stops during shuttling. All gate voltages oscillate with the conveyor frequency $f_\mathrm{conv}$, and the constant offsets $vG^\mathrm{offset}$ and amplitudes $vG^\mathrm{amp}$ are unique to each individually-controlled gate. The set of parameters $\{f_\mathrm{conv},vG^\mathrm{offset}, vG^\mathrm{amp}\}$ constitutes a conveyor definition. The pulses are applied in addition to the constant DC bias voltages $vG^\mathrm{DC}$.

All shuttling experiments begin with the ancilla A1 located below P2. A linear voltage ramp to $\{vG(0)\}$ pulls the charge below $\mathrm{C}0$, and the traveling-wave potential minimum progresses by four gate lengths for every conveyor cycle of duration $t_\mathrm{conv} = 1/f_\mathrm{conv}$. The ancilla is therefore localized adjacent to bus stops 1-4 at $t/t_\mathrm{conv} = 0.25, 1.25, 2.25, 3.25$ respectively. When the conveyor voltages are held constant at a fixed time, additional gate voltage pulses (e.g. for activating an exchange interaction) are applied in linear combination with the paused conveyor. The direction of the traveling-wave potential is reversed by flipping the sign of $f_\mathrm{conv}$ and adding a time-dependent offset to ensure there is no sudden discontinuity in the conveyor. The charge returns from below $\mathrm{C}0$ to below $\mathrm{P}2$ by linearly ramping all conveyor voltages to zero.

Two conveyor definitions are used during operation of the array. The first is a coarse-tuned conveyor that is used to shuttle uninitialized spins past unpopulated bus stops during loading and unloading. This is achieved by manually tuning $\{vG^\mathrm{amp}\}$ and, if necessary, the DC gate voltages, to ensure good charge shuttling fidelity. Coherent spin shuttling fidelity is unimportant for this conveyor definition.

The second conveyor definition is used when the bus stops are loaded with single electrons, and therefore the electrostatic landscape of the bus is different than during loading (which is ordered from bus stop 4 to bus stop 1 to avoid this issue). Here, coherent spin transfer is relevant, and we make use of CMA-ES optimization to fine tune $\{vG^\mathrm{offset}, vG^\mathrm{amp}\}$. As a cost function, we use an echo-type experiment where the spin is shuttled 10 times back-and-forth in total from $\mathrm{P}2$ to $\mathrm{C}13$ \cite{katiraeefar2025unifiedevolutionaryoptimizationhighfidelity}. The spin visibility serves as a proxy for the dephasing experienced during shuttling making it an easy-to-evaluate cost function for optimization.

The Larmor frequencies along the optimized conveyor as shown in Fig.~\ref{fig1}c are extracted by preparing the shuttled spin in superposition and observing the precession frequency after transporting the spin to various points along the array.

\begin{figure*}
    \centering
    \includegraphics[width=\textwidth]{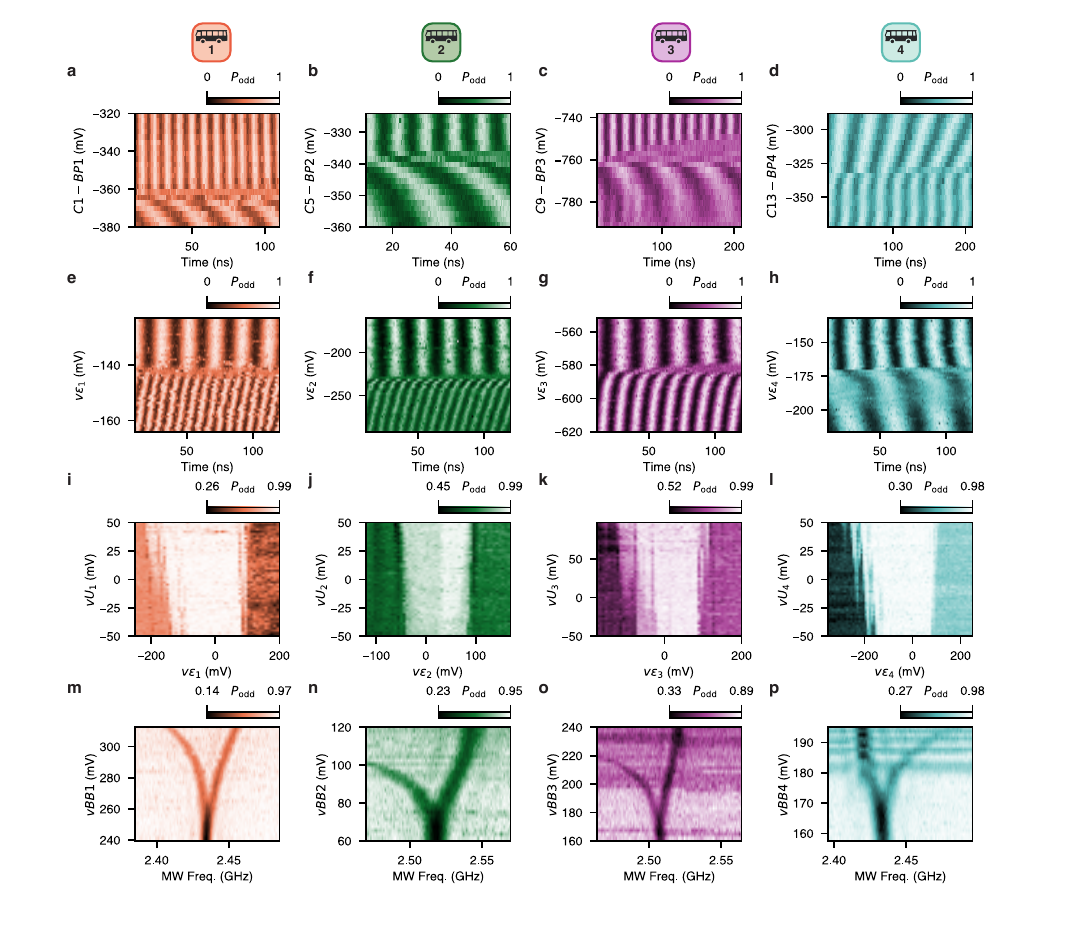}
    \caption{\textbf{a-d} First signatures of remote single-electron charge transitions measured within the four bus stops when exploring the empty array by shuttling a spin in superposition before gate virtualization or tunnel coupling tuning had taken place. \textbf{e-h} Remote single-electron charge transitions measured within the four bus stops after virtualizing control of each effective double-dot system and increasing the tunnel coupling. A fixed ramp time of \SI{40}{\nano\second} was used, therefore the tunneled spin picks up a phase that depends on the magnitude of the detuning pulse. The Larmor precession frequency in each dot is effectively constant. \textbf{i-l} Remote two-electron charge stability diagrams measured for all four bus stops when all other bus stops are populated with a single-electron. Multiple features parallel to the charge transition line likely indicate tunneling to excited orbital states. \textbf{m-p} Exchange splitting measured with EDSR for all four bus stops when all other bus stops are populated with a single electron. The four bus stops are color-coded for clarity, but the measured signal is the probability of an odd-parity PSB outcome $P_\mathrm{odd}$ in all cases.}
    \label{fig:bus_stop_calibrations}
\end{figure*}

\begin{figure*}
    \centering
    \includegraphics[width=\textwidth]{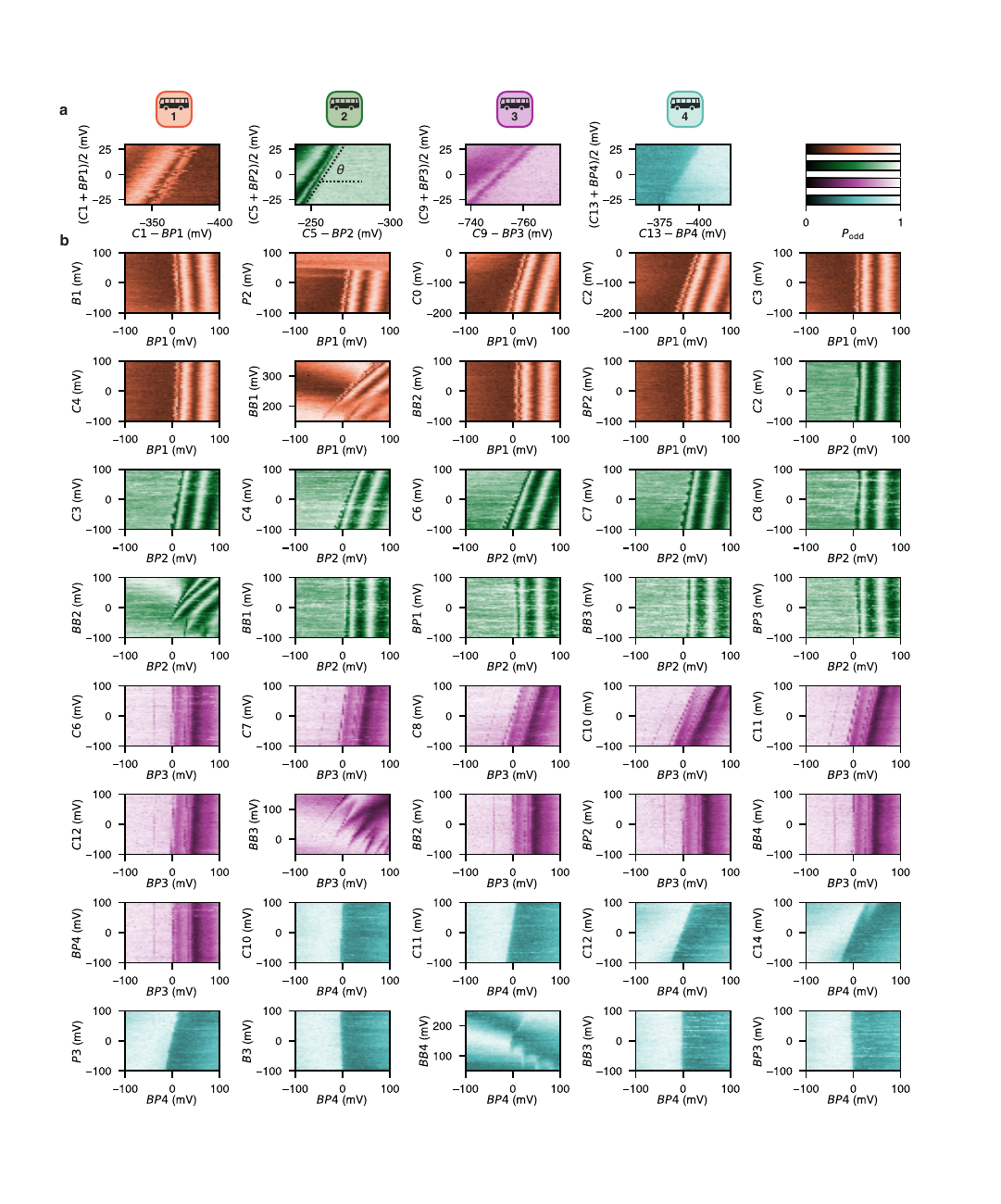}
    \caption{\textbf{a} Single-electron charge stability diagrams measured via the phase pick-up of a tunneled spin in superposition for each of the four bus stops. The four bus stops are color-coded for clarity, but the measured signal is the probability of an odd-parity PSB outcome $P_\mathrm{odd}$ in all cases. The angle $\theta$, which allows for calibration of the virtual detuning and average potential as described in Sec.~\ref{supp:1etuning}, is labeled. \textbf{b} Analogous gate-gate virtualization measurements for all gates $G$ in the vicinity of each bus stop. The extracted slopes $\mathrm{d}BPi/\mathrm{d}G$ from all scans are represented as a heat map in Fig.~\ref{fig2}g.}
    \label{fig:sparse_crosstalk}
\end{figure*}

\section{Single-spin remote tuning}
\label{supp:1etuning}

After confirming the operation of the shuttling bus, the loading protocol of the bus stops can be tuned. For this, we leverage the availability of EDSR throughout the bus to perform Ramsey experiments with respect to various rotating frames, but we note that the strategy we use is compatible with not having single-spin control throughout the array. The minimum requirement is the ability to prepare a spin in a superposition state and shuttle it coherently.

During initial tuning, no gate virtualization has taken place beyond the readout zone, and the ability to load the bus stops is probed by applying a physical detuning pulse $\epsilon_i = Cj-BPi$ where Cj is the conveyor gate adjacent to bus stop $i$ ($j = 1,5,9,13$ for bus stops 1-4 respectively). Fig.~\ref{fig:bus_stop_calibrations}a-d shows the first convincing experimental signatures of bus stop loading before any virtualization takes place. To verify that the bus stop is the most likely destination for the tunneled electron, we repeat the loading experiments while pulsing negatively on the surrounding conveyor gates. This provides evidence that tunneling is not taking place within the shuttling bus.

The Larmor frequency difference between the spin when localized in the bus and in the bus stop encodes information about its position in the phase of the spin state. For each bus stop, we select a free induction time on the order of tens of nanoseconds to optimize contrast between the two cases, and we use this signal to virtualize control of the effective double-dot system as well as the surrounding conveyor gates. First, we focus on the gate voltage subspace consisting of the physical conveyor gate voltage $C_j$ and physical bus stop plunger gate voltage $BPi$. In the case of bus stop 1, these gates are already virtualized with respect to the readout zone, but not to each other, and the following procedure is the same regardless of any pre-existing virtualization.

We define unvirtualized detuning and average potential parameters $\epsilon_i, U_i$ that relate to voltages $Cj, BPi$ as $\left(\epsilon_i, U_i\right)^T = V\left(Cj, BPi\right)^T$ where:

\begin{equation}
    V = \begin{pmatrix} 1 & -1 \\ 0.5 & 0.5 \end{pmatrix}.
\end{equation}

We want to identify corresponding virtual gates $vCj$ and $vBPi$, where $\left(vCj, vBPi\right)^T = M_i\left(Cj, BPi\right)^T$, such that the virtual detuning and average potential, defined as $\left(v\epsilon_i, vU_i\right)^T = V\left(vCj, vBPi\right)^T$, relate to the double-dot chemical potentials as sketched in Fig.~\ref{fig2}a. Explicitly, this means $v\epsilon_i\propto \mu_\mathrm{Cj}-\mu_\mathrm{BPi}$ and $vU_i\propto (\mu_\mathrm{Cj}+\mu_\mathrm{BPi})/2$ where $\mu_\mathrm{Cj}$ and $\mu_\mathrm{BPi}$ are the chemical potentials of the dots formed below gates $\mathrm{Cj}$ and $\mathrm{BPi}$ respectively. An interdot charge transition to bus stop $i$ should therefore only be modulated by $v\epsilon_i$.

The interdot charge transitions measured when sweeping the physical detuning and average potential parameters (see Fig.~\ref{fig:sparse_crosstalk}a) show a finite (positive) slope $\mathrm{d}\epsilon_i/\mathrm{d}U_i = \tan\theta_i$ from which it follows that $\left(v\epsilon_i, vU_i\right)^T = R(\theta_i)\left(\epsilon_i, U_i\right)^T$ where $R(\theta_i)$ is a counterclockwise rotation about the origin $\left(\epsilon_i = v\epsilon_i = 0, U_i = vU_i = 0\right)$ by an angle $\theta_i$ such that $\tan(\theta_i) = d\epsilon_\mathrm{BS,i}/dU_\mathrm{BS,i}$. We therefore have $M_i = V^{-1}R(\theta_i)V$. Applying the elements of $M_i$ to the total virtual gate matrix permits direct control over the virtual detuning and average potential.

To virtualize any other gate $G$ with respect to each bus stop plunger $\mathrm{BP}i$, the same experiment definition is used where the slope $\mathrm{d}BPi/\mathrm{d}G$ of the interdot transition informs both the degree of influence of the gate voltage $G$ on the chemical potential below gate $\mathrm{BP}i$ and therefore the virtual gate definition $\mathrm{vG}$ required to offset the effect (see Fig.~\ref{fig:sparse_crosstalk}b). The slope $\mathrm{d}BPi/\mathrm{d}Cj=(2-\tan\theta_i)/(2+\tan\theta_i)$. This provides all slopes plotted in the crosstalk heatmap of Fig.~\ref{fig2}g.

Fig.~\ref{fig:bus_stop_calibrations}e-f shows the loading experiments repeated after gate virtualization has taken place. In certain cases, the tunnel coupling between the bus and bus stop dots is pulsed during tunneling to smoothen the transition compared to the preliminary loading. The two-electron charge stability diagrams reconstructed from the shuttled spin polarization in Fig.~\ref{fig:bus_stop_calibrations}i-l also show charge transitions that are effectively orthogonal to the virtualized detuning axis, confirming the validity of this approach.

\begin{figure}
    \centering
    \includegraphics[width=\columnwidth]{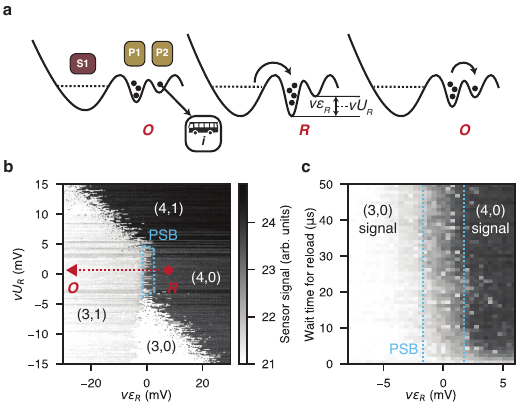}
    \caption{\textbf{a} Illustration of the reload protocol. After shuttling the ancilla spin to a bus stop, the charge state of the readout pair is blockaded in the metastable (3,0) charge state at the operation point \textit{O}. When the detuning is pulsed to the reload point \textit{R}, the blockade is lifted and a fourth electron quickly tunnels from S1 to the readout ancilla dot. The double-dot system returns to the (3,1) charge state upon ramping the detuning back to the point \textit{O}. \textbf{b} rf-reflectometry measurement of the equilibrium charge stability diagram of the readout pair. The PSB region is highlighted by the blue dashed lines. The detuning axis followed to go from the operation point \textit{O} to the reload point \textit{R} is shown by the red dashed line. \textbf{c} Averaged rf-reflectrometry time-traces after waiting at different reload points along the detuning axis of \textbf{b} after the system has been initialized in the metastable (3,0) charge state. Between \textit{O} and the PSB region, blockade persists for at least \SI{20}{\milli\second}. Beyond the PSB region, the reload time scale is on the order of \SI{10}{\micro\second}. The colorbar is identical to that of \textbf{b}.}
    \label{fig:reload}
\end{figure}

\section{Spin Reloading}
\label{supp:reload}

Fig.~\ref{fig:reload}a illustrates the procedure used for reloading a new spin into the array after the original ancilla has been shuttled to occupy a bus stop. The sensor S1 doubles as an electron reservoir from which qubits can be loaded on-demand. It has a small but finite tunnel coupling to the dot below P1 where the readout ancilla R1 remains and a weaker coupling to the dot below P2 where the ancilla A1 is initialized as can be seen in the measured charge stability diagram of Fig.~\ref{fig:reload}b.

After shuttling the ancilla, the voltage configuration in the readout zone remains at the operation point $O$ in the quasi-equilibrium (3,0) charge state. As the direct tunnel rate to (3,1) is very slow, this charge configuration persists for at least \SI{20}{\milli\second} and is limited by the bias-tee charging time of the sample PCB. No unwanted electrons tunnel into the empty dot during the experiments. To reload on-demand, the virtual detuning of the readout pair $v\epsilon_\mathrm{R}$ is pulsed to the reload point $R$. The rate at which a new charge enters is evident from Fig.~\ref{fig:reload}c. Pulsing beyond the PSB region of the interdot transition allows for a relatively fast loading to the (4,0) state in less than \SI{10}{\micro\second}, and ramping the detuning back to $O$ returns the system to the equilibrium (3,1) charge state and restores the ancilla spin. No additional barrier pulse (i.e. on gate B0) is used to modulate the tunnel rate, and including one could increase the loading speed further. This reloading procedure repeats until all bus stops are occupied and the final ancilla spin is reloaded, after which the initialization protocol may begin.

The bus stop spins are unloaded in one of two ways. First, they may be shuttled back to the readout zone in a sequence inverse to the loading protocol. In this case, the five-electron charge state at the readout zone quickly equilibrates to the (3,1) charge state in about \SI{1}{\micro\second}, and no detuning pulse is necessary. Alternatively, we can simply ramp all pulsed voltages back to their DC condition without any traveling-wave potential used. Bias-tee compensation pulses are applied over a period of several tens of microseconds before the next experimental shot begins, and the free spins vacate the bus stops within this time scale, presumably to one of the 2DEG reservoirs on either side of sparse array. We observe no adverse consequences when unloading spins in this manner, and this approach is used for most experiments presented in this work.

The necessity for unloading and subsequent loading is set by the finite bias-tee time and amplitude range of the AWG modules. In future work, loading bus stops or other remote quantum dots with DC voltages would eliminate the need for spin reloading altogether.

\section{QND Measurement}
\label{supp:QNDmeasurement}

All qubits in the effective five-qubit processor can be measured in the computational basis as illustrated in Fig.~\ref{fig3}d by a sequence of parity-mode PSB measurements $m_0, m_1, m_2, \dots$. We use the convention that each projective measurement ideally yields $m_k=+1$ when an odd-parity spin state $\ket{R_1}\otimes\ket{A1}$ is present and $m_k=-1$ when an even-parity spin state is present.

\begin{figure}
    \centering
    \includegraphics[width=\columnwidth]{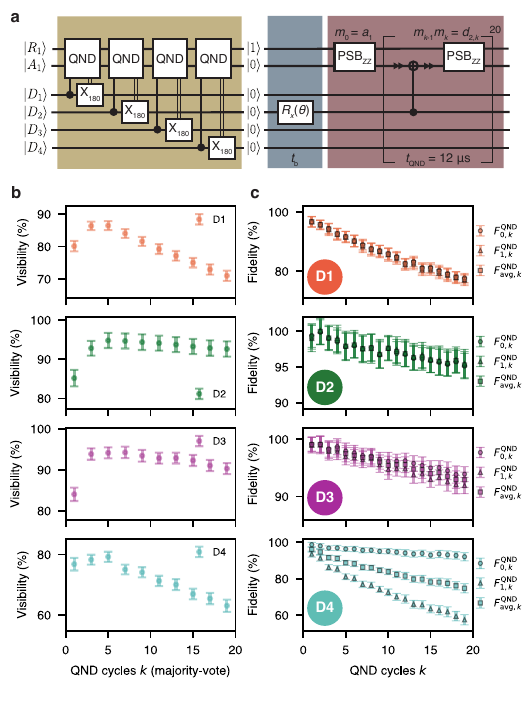}
    \caption{\textbf{a} Example of a repeated QND measurement circuit for D2. After the loading and initialization of all qubits, a resonant microwave burst is applied to one of the data qubits for a time $t_\mathrm{b}$. This is followed by 20 successive QND measurement cycles on the data qubit. The single-shot computational-basis measurement for the $k$-th cycle is given by $d_{i,k}=m_{k-1}m_k$. Each QND measurement cycle requires about \SI{12}{\micro\second}. \textbf{b} Visibility of data qubit Rabi oscillations as a function of the number of successive QND measurement cycles used to conduct a majority-vote. \textbf{c} QND fidelities calculated as a function of the number of successive QND cycles.}
    \label{fig:qnd_analysis}
\end{figure}

The ancilla is always measured first. By assuming R1 remains in the prepared eigenstate $\ket{1}$ throughout all experiments, $m_0=a_1$ gives the computational-basis measurement outcome of qubit A1. To perform a QND measurement on a data qubit D$i$, A1 is shuttled adjacent to the relevant bus stop and a calibrated CROT will flip the state of A1 conditional on the data qubit being in the $\ket{1}$ state (see Sec.~\ref{supp:universalcontrol}). The next measurement $m_k$ indicates whether an ancilla spin flip took place by considering the previous measurement $m_{k-1}$. The $k$-th computational-basis readout $d_{i,k}$ of D$i$ is therefore given by the product $m_{k-1}m_k$.

Unlike for the ancilla measurement, repeated QND measurements can take place on the data qubits to improve the readout fidelity as indicated in Fig.~\ref{fig3}a. To characterize the quality of the repeated QND readout, we perform Rabi oscillation measurements with 20 sequential QND readouts on the four data qubits individually. Fig.~\ref{fig:qnd_analysis}a gives an exemplary circuit for D2. The QND single-shot outcomes $\{d_i\} = \{d_{i,1},d_{i,2},\dots d_{i,20}\}$ are used to infer the data qubit D$i$ $\ket{1}$ state probability $P_1(t_\mathrm{b})$ through a majority-vote scheme. The resulting oscillation is fit to $P_{1}(t_{\mathrm{b}}) = B - A \cos(2\pi f_{\mathrm{R}} t_{\mathrm{b}}) \exp( -t_{\mathrm{b}}/T_{2}^{\mathrm{R}})$, where $t_{\mathrm{b}}$ is the MW burst time, $f_{\mathrm{R}}$ is the Rabi frequency, and $T_{2}^{\mathrm{R}}$ is the Rabi oscillation decay time. Fig.~\ref{fig:qnd_analysis}b shows the visibility, given by $2A$, of the Rabi oscillations as a function of the number of QND readout cycles used in the majority-vote. We use up to five QND repetitions in our experiments, as this yields the highest visibility for all data qubits.

We also characterize the QND fidelity of our readout, which quantifies the ability of the QND measurement to preserve the data qubit $\ket{0}$ and $\ket{1}$ states. We again utilize the 20 repetitive QND readouts but calculate the Rabi oscillation from each individual outcome $d_{i,k}$. This results in Rabi oscillations with decaying amplitudes and offsets as the data qubit state is disturbed by finite CROT fidelity and spin relaxation. The individual Rabi oscillations $P_{1,k}(t_\mathrm{b})$ are fitted to obtain the amplitude $A_{k}$ and offset $B_{k}$ for the $k$-th QND measurement repetition. The decay of these parameters is then fitted using the model \cite{Kobayashi_feedback_2023}:
\begin{align}
    A_{k} &= A_{0}\gamma^{k}, \\
    B_{k} &= \left( B_{0} - \frac{1-\gamma_{0}}{1-\gamma}\right) \gamma^{k} + \frac{1-\gamma_{0}}{1-\gamma},
\end{align}
where $\gamma_{0}=\exp(-t_{\mathrm{QND}}/T_{1}^{(0)})$, $\gamma_{1}=\exp(-t_{\mathrm{QND}}/T_{1}^{(1)})$, and $\gamma = \gamma_{0}+\gamma_{1}-1$. $t_\mathrm{QND}$ is the time required for a cycle of QND measurement, which in our case is about \SI{12}{\micro\second} for all four data qubits, $T_1^{(0)}$ is the spin-flip time of the data qubit $\ket{0}$ state, and $T_1^{(1)}$ is the spin-flip time of the data qubit $\ket{1}$ state. We observed a decay in $A_{k}$ for all data qubits. However, only data qubit D4 shows a decay in $B_{k}$; the remaining data qubits show no visible decay over the 20 QND repetitions. We conclude that the QND readouts of D1, D2, and D3 are mainly limited by unintentional data qubit spin flips due to the CROT infidelity, while the QND readout of D4 is affected by both the CROT infidelity and instrinsic spin relaxation. We are unsure of the reason for this unique spin relaxation effect. One possible explanation could be that the valley splitting of the dot containing D4 is nearly degenerate with the Zeeman splitting, as spin relaxation would be enhanced. This potential issue may be bypassed by changing the magnetic field to lift the degeneracy.

The fitted parameters $A_{k}$ and $B_{k}$ are related to the QND fidelities by the following relation \cite{Kobayashi_feedback_2023}:
\begin{align}
    P_{1,k}(t_{\mathrm{b}}) = (1-F_{0,k}^{\mathrm{QND}}) (1-P_{1,0}(t_{\mathrm{b}})) + F_{1,k}^{\mathrm{QND}} P_{1,0}(t_{\mathrm{b}}),
\end{align}
which translates to:
\begin{align}
    F_{0,k}^{\mathrm{QND}} &= 1 - B_{k} + \frac{B_{0}}{A_{0}} A_{k}, \\
    F_{1,k}^{\mathrm{QND}} &= B_{k} + \frac{1-B_{0}}{A_{0}} A_{k}. 
\end{align}
Figure~\ref{fig:qnd_analysis}c shows the extracted QND fidelities $F_{0,k}^{\mathrm{QND}}$ and $F_{1,k}^{\mathrm{QND}}$, along with the average fidelity $F_{\mathrm{avg},k}^{\mathrm{QND}} = (F_{0,k}^{\mathrm{QND}} + F_{1,k}^{\mathrm{QND}})/2$. For one QND repetition, the $F_{\mathrm{avg},1}^\mathrm{QND}$ for D1-D4 are 96.7(1.8)\%, 99.2(1.7)\%, 99.1(1.3)\%, and 96.1(1.4)\% respectively.

\section{Universal spin control}
\label{supp:universalcontrol}

All primitive qubit operations can be derived from the Heisenberg Hamiltonian describing the sparse spin array. It can be expressed in terms of the Loss-DiVincenzo qubit operators as:

\begin{equation}
    H = \sum_{ij}hJ_{ij}\left(\frac{\vec{\sigma}_i\cdot\vec{\sigma}_j}{4}-\frac{1}{4}\right) - \sum_i\frac{1}{2}g\mu_B\vec{B}_i\cdot\vec{\sigma}_i,
\end{equation}

\noindent where $\vec{\sigma}_i = \left(X, Y, Z\right)^T$ is the vector of Pauli matrices acting on qubit $i$, $\vec{B}_i$ is the magnetic field vector at the location of qubit $i$, $h$ is the Planck constant, $g\approx2$ is the electron spin $g$-factor in silicon, and $\mu_B$ is the Bohr magneton. $J_{ij}$ gives the magnitude of exchange interaction between spins. It is controllable via baseband pulses on the electrostatic gates and is effectively zero for non-adjacent spins.

When $J=0$, single-qubit control is achieved via EDSR where an oscillating electric field $E_\mathrm{ac}(t)\cos(2\pi f_\mathrm{MW}t+\phi)$ with time-varying amplitude $E_\mathrm{ac}(t)$, frequency $f_\mathrm{MW}$ and phase $\phi$ couples to the spin via a transverse gradient $b_t$ originating from the micromagnet stray field. The single spin therefore experiences an effective magnetic field $\vec{B}(t) = \left(h f_\mathrm{R}\cos(2\pi f_\mathrm{MW}t+\phi), 0, h f_\mathrm{L}\right)^T/g\mu_B$ where $f_\mathrm{L}$ is the Larmor frequency set by the total magnetic field at the qubit location and $f_\mathrm{R}(t) = g\mu_B b_t e E_\mathrm{ac}(t)a_0^2/2hE_\mathrm{orb}$ is the on-resonance Rabi frequency, $e$ is the electron charge, $a_0$ is the Fock-Darwin length scale of the quantum dot and $E_\mathrm{orb}$ is the orbital energy scale of the quantum dot.

In the rotating frame, the single-qubit Hamiltonian after application of the rotating wave approximation is:

\begin{equation}
    H_\mathrm{EDSR} = \frac{h(f_\mathrm{MW}-f_\mathrm{L})}{2}Z + \frac{hf_\mathrm{R}(t)}{2}\left(\cos\phi X-\sin\phi Y\right),
\end{equation}

\noindent When driven on-resonance such that $f_\mathrm{MW}=f_\mathrm{L}$, the single-qubit unitary evolution is given by

\begin{equation}
    \label{eq:1Qunitary}
    U_{1Q} = \exp\left(-i 2\pi f_\mathrm{R}(t)t\left(\cos\phi,-\sin\phi,0\right)^T\cdot\vec{\sigma}/2\right),
\end{equation}

\noindent which is a rotation $R_{\hat{n}}(\theta)$ about an axis $\hat{n}=\left(\cos\phi,-\sin\phi,0\right)^T$ by an angle $\theta = 2\pi f_\mathrm{R}(t)t$. All single-qubit gates are derived from a single definition for an $X_{90} = \exp\left(-i\pi X/4\right)$ operation which consists of a microwave burst of duration $t_{90}$. Rotations around other axes are implemented by changing the phase $\phi$ of the microwave burst, and arbitrary rotations $R_z(\theta)$ about the $z$-axis of the Bloch sphere are similarly implemented by a phase update of the subsequent microwave bursts. $X_{180}$ rotations are implemented with two concatenated $X_{90}$ pulse definitions. The use of a unique single-gate primitive streamlines the calibration of crosstalk in the multi-qubit Hilbert space (see Sec.~\ref{supp:gatecalibration}).

As all microwaves are delivered via a single antenna, we rely on spectral separation to address individual spins. To limit errors due to off-resonant coherent driving, we use a Hamming window for the microwave pulse amplitude $E_\mathrm{ac}(t)$ to limit the bandwidth of each microwave pulse to approximately $1/t_{90}$. For pulse durations of about \SI{250}{\nano\second}, the pulses therefore have a resolution of about \SI{10}{\mega\hertz}. Remaining phase pickup due to the ac Stark shift and heating effects is calibrated experimentally and accounted for in the form of virtual phase updates.

A CROT operation is used for initialization and readout by applying a resonant pulse while the exchange $J_{ij}$ between the shuttled ancilla and a data qubit is finite. In the limit where the exchange is much smaller than the Larmor frequency difference between spins ($J_{ij}\ll \left|f_{\mathrm{L},i}-f_{\mathrm{L},j}\right|$), it is appropriate to approximate the resonance frequency of qubit $i$ as conditional on the state of qubit $j$ with a separation of $J_{ij}$. Qubit $i$ can therefore be conditionally rotated depending on the state of qubit $j$ via EDSR. The exchange $J_{ij}(t)$ is ramped on and off adiabatically with a Tukey pulse shape, and a microwave burst with a rectangular window is applied for a duration $t_{180}$ required for a complete flip of qubit $i$. The synchronization condition $1/2t_{180} = J_{ij}/\sqrt{4n^2-1}$ with integer $n$ is used to drive a $2\pi$ rotation in the subspace of the undesired transition \cite{Russ_2018}. The operation therefore performs the mapping $\{\ket{00}\rightarrow\ket{00}, \ket{01}\rightarrow\ket{01}, \ket{10}\rightarrow\ket{11}, \ket{11}\rightarrow\ket{10}\}$ which is suitable for a projective QND measurement of the control qubit in the computational basis. Although the CROT may form a universal entangling gate, we do not use the CROT as a coherent two-qubit operation and therefore do not track the single-qubit phases picked up during the operation. We also neglect the conditional phases in the CROT as they are unimportant for initialization and readout.

Coherent two-qubit operations are implemented by modulating the exchange $J_{ij}$ adiabatically without any additional microwave burst. This ideally results in a unitary evolution $U_\mathrm{2Q} = \mathrm{diag}\left(1, e^{-i\phi_{01}}, e^{-i\phi_{10}},e^{-i\phi_{11}}\right)$. $U_\mathrm{2Q}$ can be transformed into a controlled-phase gate via commuting virtual single-qubit $z$-rotations such that $R_z(\phi_{10})\otimes R_z(\phi_{01})U_\mathrm{2Q} = \mathrm{diag}\left(1,1,1,e^{i(\phi_{01}+\phi_{10}-\phi_{11})}\right)$. A controlled-$S$ gate between qubits $i$ and $j$ is achieved when $J_{ij}(t)$ is modulated such that $\phi_{01}+\phi_{10}-\phi_{11}=\pi/2$. The duration for the gate is bounded by $t_{CS}>1/4J_{ij,\mathrm{max}}$, where $J_{ij,\mathrm{max}}$ is the maximum exchange strength during the operation, and is necessarily longer to maintain adiabaticity.

The $CS$ gate is used as the two-qubit entangling primitive as illustrated in Fig.~\ref{fig4}a where the single-qubit $z$-rotations are obviated by the refocusing pulses. The multi-qubit decoupled CZ operation can be expressed in terms of controlled-phase gates $CZ_{00}^i$ acting on the ancilla qubit and data qubit $i$, where $CZ_{xy}\ket{m,n} = \left(-1\right)^{\delta(x,m)\delta(y,n)}\ket{m,n}$. The circuit of Fig.~\ref{fig4}a implements the unitary $X_{180}^{\otimes (w+1)}\Pi_{i=1}^w CZ_{00}^{i}$ acting on the ancilla qubit and $w$ data qubits. With a universal single-qubit gate set, this operation is sufficient to apply any Pauli operation on the data qubits conditional on the state of the ancilla qubit.

\begin{figure*}
    \centering
    \includegraphics[width=\textwidth]{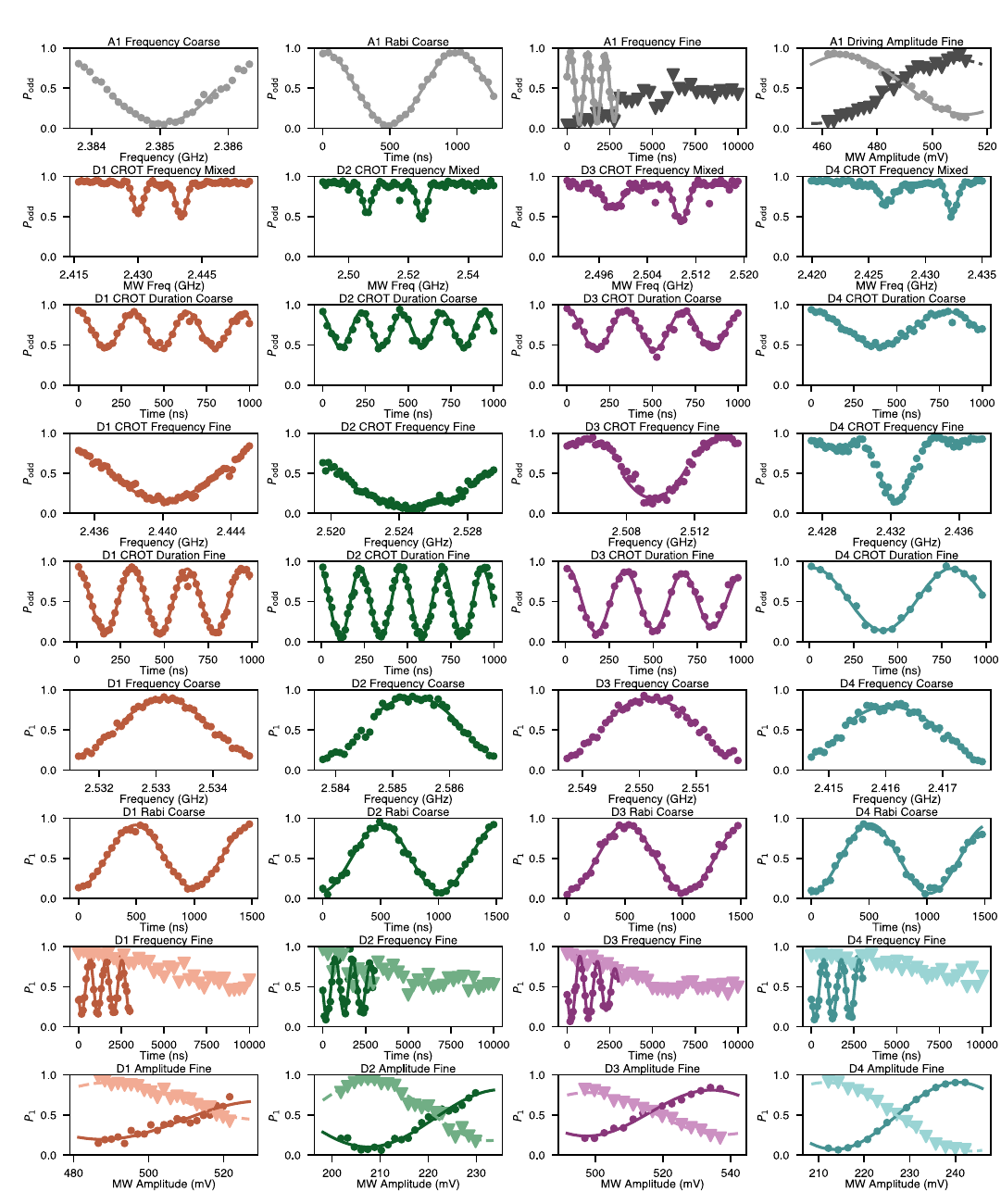}
    \caption{A calibration report for all resonant operations. The order of the experiments is from left-to-right and top-to-bottom. For fine frequency experiments, circles indicate a scan with a virtual detuning, and triangles indicate a verification scan. For fine driving amplitude experiments, circles and triangles indicate whether a final $X_{-90}$ or $X_{90}$ gate was applied. Ancilla measurements are reported as the probability $P_\mathrm{odd}$ of an odd-parity PSB outcome. Data qubit QND measurements are reported as the probability $P_1$ of inferring the $\ket{1}$ state.}
    \label{fig:report}
\end{figure*}

\begin{figure}
    \centering
    \includegraphics[width=\linewidth]{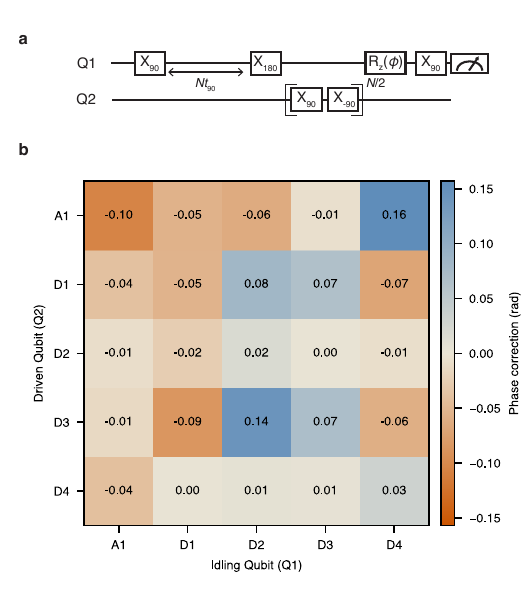}
    \caption{\textbf{a} Quantum circuit used to perform a Hahn-echo-type experiment and characterize the single-qubit phase crosstalk accrued by Q1 while performing single-qubit operations on Q2. \textbf{b} An example of the crosstalk phase corrections that are applied along with each corresponding $X_{90}$ gate while operating the device as a five-qubit processor.}
    \label{fig:phase_crosstalk}
\end{figure}

\begin{figure*}
    \centering
    \includegraphics[width=\textwidth]{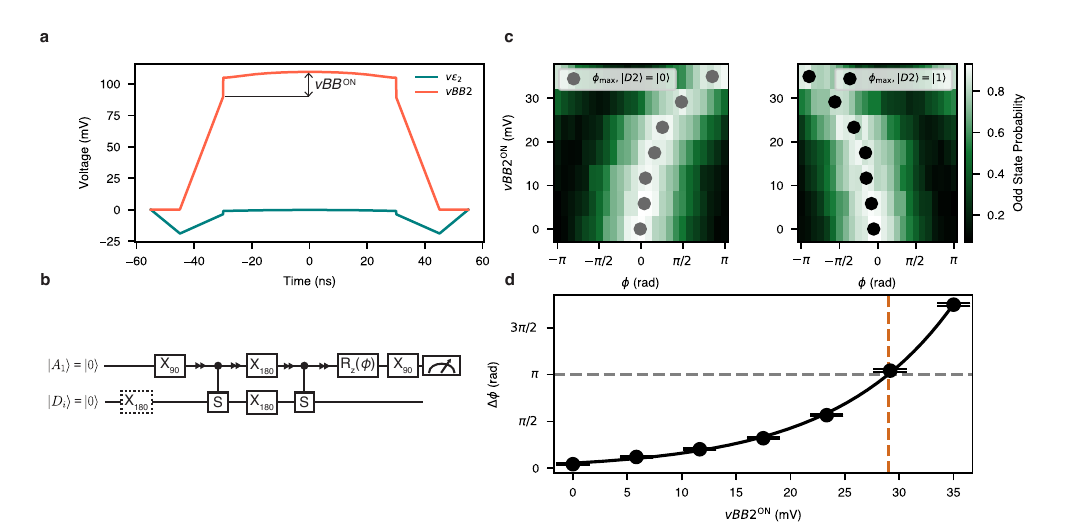}
    \caption{\textbf{a} An example of a Hamming-window exchange pulse for implementing a $CS$ gate between A1 and D2. \textbf{b} Quantum circuit used to fine-tune the maximum amplitude of the Hamming pulse. \textbf{c} The phase pickup of the ancilla qubit for different maximum Hamming pulse amplitudes for both computational-basis preparations of the data qubit D2. The total gate time is fixed. The markers indicate the relative shift of the curves due to conditional phase pickup. \textbf{d} The difference in the conditional phase picked up by the ancilla is fit and we select the barrier amplitude that yields a total phase difference of $\pi$ radians from the two successive $CS$ gates.}
    \label{fig:DCZ_tuning}
\end{figure*}

\section{Gate calibration protocols}
\label{supp:gatecalibration}

Periodic calibration of the five-qubit processor is necessary to compensate for slow drifts in the solid-state environment in order to maintain the gate fidelities required for the results presented in this work. We use a semi-automated calibration procedure that efficiently updates the control parameters for resonant gates (Fig.~\ref{fig:report}), crosstalk compensation (Fig.~\ref{fig:phase_crosstalk}), and adiabatic two-qubit gates (Fig.~\ref{fig:DCZ_tuning}). The following protocols may be carried out once all bus stops can be populated, the ancilla can be shuttled with reasonable fidelity (specifically, the visibility of QND measurement of the data qubits will be bounded by how well spin polarization is preserved during shuttling), and all four exchange interactions are tunable over a workable range (e.g. \SI{100}{\kilo\hertz} < $J$ < \SI{10}{\mega\hertz}).

\subsection{Single-qubit gate and CROT calibration}
\label{supp:gatecalibration_singleQandCROT}

Each resonant gate calibration begins with a coarse calibration followed by a finer one. We coarsely calibrate the Larmor frequency of the ancilla qubit A1 using microwave spectroscopy. The dip in the measured odd-state return probability is isolated and the resonance frequency is estimated via a quadratic fit. We then drive A1 with a microwave burst shaped by Hamming window of varying duration and we fit the Rabi oscillations to extract the current Rabi frequency \(f_\mathrm{R}^{\mathrm{fit}}\). Since we aim for \(f_{\mathrm{R}}^{\mathrm{targ}} = 1~\mathrm{MHz}\), or $t_{90}=250~\mathrm{ns}$, we rescale the driving amplitude by \(f_{\mathrm{R}}^{\mathrm{targ}}/f_\mathrm{R}^{\mathrm{fit}}\).

A finer calibration is then performed using a Ramsey experiment, with the rotating frame virtually detuned by \SI{1}{\mega\hertz} from the coarsely-calibrated Larmor frequency. Fitting the resulting Ramsey oscillations yields a frequency correction. Finally, we fine-tune the \SI{250}{\nano\second} microwave burst amplitudes. The procedure applies multiple \(2\pi\) qubit rotations (typically 4), each decomposed into four repeated \(\mathrm{X_{90}}\) gates, to amplify any systematic over-rotation. The sequence is terminated with either an \(\mathrm{{X}_{90}}\) or a \(\mathrm{{X}_{-90}}\) gate to prepare the qubit in a superposition state. By sweeping the pulse amplitude, we select the value that yields a 50\% odd-parity measurement probability for both final states (corresponding to $\langle Z\rangle = 0$). The last two calibration steps are often iterated until they converge.

The CROT is calibrated using a four-step procedure, beginning with a coarse frequency estimation obtained by fitting a double-Gaussian function to an exchange spectroscopy trace acquired with the relevant data qubit in a mixed state and the exchange activated. This is possible before resonant control of the data qubits has been calibrated. For all CROTs, the higher-frequency branch is selected to drive the ancilla conditional on the data qubits occupying the $\ket{\uparrow}\equiv\ket{1}$ state. The Rabi frequency of the conditional rotations is used to calibrate the driving amplitude required to meet the synchronization condition.

The coarse CROT calibrations are sufficient to initialize the data qubits using QND measurement and postselection of the desired CROT outcome, and resonant control of D1-D4 may then be calibrated analogously to A1. A finer CROT calibration is then performed by initializing the data qubits to the $\ket{1}$ state. The same calibrations are repeated, yielding more accurate estimates of both the conditional resonance frequency and the required drive amplitude due to the improved trace visibility.

A finer calibration of the CROTs, along with resonant control of the data qubits, allows for higher visibility of the data qubits by using the full QND measurement framework presented in Fig.~\ref{fig3}. Consequently, resonant control of the data qubits can be calibrated more precisely, and the process can be iterated until satisfactory convergence of the parameters has been achieved.

\subsection{Single-qubit phase correction}
\label{supp:phase_crosstalk}
As all resonant control signals are applied to the same screening gate, operating the device as a five-qubit processor requires mitigating crosstalk effects when driving different qubits that may arise due to a combination of the AC Stark effect, device heating, and induced shifts in the stray magnetic field gradient. As the addressability gradient and pulse-shaping limit off-resonant rotations given by Eq.~\ref{eq:1Qunitary}, this crosstalk predominantly manifests as phase pickup, which can be compensated for by virtual $R_z(\phi)$ gates.

To characterize these phases and determine the corresponding per-gate phase corrections, we perform a series of Hahn-echo-like experiments as shown in Fig.~\ref{fig:phase_crosstalk}a in which the target qubit remains idle while repeated sequences of \(\mathrm{X_{90}}\) and \(\mathrm{X_{-90}}\) gates are applied to each of the other qubits. We limit the number of applied gates to a maximum of \(\mathrm{N = 4}\) to remain in the linear regime, allowing us to extract the per-gate accumulated phase via a linear fit.

These induced phases can be compared with the predicted AC Stark shifts, showing partial qualitative agreement. However, the presence of self-induced phase accumulation (non-zero diagonal elements) as well as magnitude and sign discrepancies with the experimental data indicate that other effects, such as heating, play a significant role.

\subsection{Adiabatic CS calibration}
\label{supp:CZgate}

We calibrate coherent two-qubit interactions between the ancilla qubit and data qubits using an approach similar to \cite{Wang_2024}. The dynamic range of the exchange interaction is probed to find a suitable empirical relation $J_i(vBBi)$ for all four bus stops. The charge-symmetry point is also identified via a fingerprint scan as an approximately linear function of the barrier voltage such that $v\epsilon_i \propto vBBi$. This allows for pulse-shaping of the exchange strength $J_i(t)$ to maximize the adiabaticity of the interaction while remaining robust to charge noise \cite{Reed_2016}.

Each effective $CS$ operation is actuated by a series of baseband pulse segments as exemplified in Fig.~\ref{fig:DCZ_tuning}a. The first segment is a fast detuning ramp to the symmetry point. The second segment is a linear ramp to a sub-threshold exchange magnitude (e.g. $J$ < \SI{1}{\mega\hertz}). The central segment is a Hamming-shaped exchange modulation characterized by a pulse duration and magnitude $vBBi^\mathrm{ON}$ which are selected to balance speed and adiabaticity. The remaining two segments are time-reversed copies of the first two to uncouple the qubits.

The pulse sequence in Fig.~\ref{fig:DCZ_tuning}b is used to fine-tune the amplitude $vBBi^\mathrm{ON}$ in a decoupled CZ sequence such that the total conditional phase picked up by A1 can be identified for both computational-basis state preparations of the relevant data qubit D$i$. This is exemplified for the two-qubit interaction at bus stop 2 in Fig.~\ref{fig:DCZ_tuning}c-d. A total conditional phase difference of $\pi$ is selected to ensure each individual $CS$ operation accrues $\pi/2$ radians of conditional phase.

\begin{figure}
    \centering
    \includegraphics[width=\columnwidth]{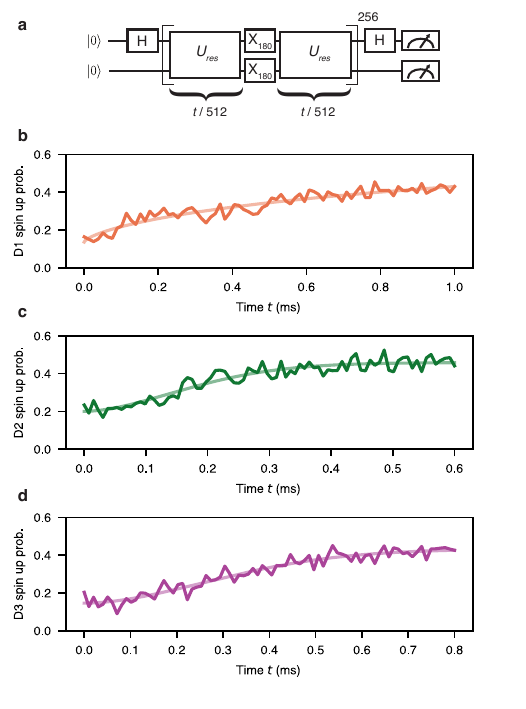}
    \caption{\textbf{a} Circuit to measure the residual exchange between two data qubits localized in bus stops. The unitary evolution under idling is Ising-like as $J_\mathrm{res}<\Delta E_\mathrm{Z}$, and $U_\mathrm{res}\approx\mathrm{diag}\left(1,e^{iJ_\mathrm{res}t/2},e^{iJ_\mathrm{res}t/2},1\right)$, therefore the oscillations of the qubit placed in superposition will have a frequency $J_\mathrm{res}/2$ as a function of the total idling duration. \textbf{b-d} Measurements of residual exchange between D1 and D2, D2 and D3, and D3 and D4 respectively. In all cases, no residual exchange on the order of kilohertz is discernible from decoherence and is therefore negligible for any experiment conducted in this work. Due to the relatively large separation and large potential barrier between data qubits, the actual residual exchange is likely much lower.}
    \label{fig:residual_exchange}
\end{figure}

\section{Single-qubit characterization}
\label{supp:1Qcharacterization}

The dephasing times measured using CPMG decoupling can be used to gain insight into the noise spectrum influencing each qubit (see Fig.~\ref{fig3}e). We assume a monotonic noise spectrum $S(\omega)=A/\omega^\alpha$ acting on each qubit such that $T_2=T_2^0N_\pi^{\alpha/(1+\alpha)}$ and $T_2^0 = (2/A)^{1/(\alpha+1)}\pi^{\alpha/(1+\alpha)}$ and observe a good fit in all cases \cite{Bylander_2011}. $T_2^*$ measurements use a total integration time of about 25 minutes.

Single-qubit gate fidelities are evaluated using randomized benchmarking. We use the primitive gate set $\{I,X_{90}, Y_{90}\}$ of only positive rotations such that a Clifford gate is composed of 3.125 primitive gates on average \cite{Lawrie_2023}. A $X_{90}$ gate duration of \SI{250}{\nano\second} is used for each primitive gate. The single-qubit Clifford gate fidelities are 98.98(3)\%, 99.930(5)\%, 99.90(1)\%, 99.87(1)\%, and 99.924(6)\% for A1, D1, D2, D3, and D4 respectively. Individual single-qubit gate fidelities in the single-qubit subspace are all well above 99.9\% for the data qubits, and slightly lower for the ancilla qubit (see Fig.~\ref{fig3}g). In the experiment, pairs of single-qubit operations were performed simultaneously when applicable. When all four bus stops were used, D1/D2 and D3/D4 were parallelized. When three bus stops were used, the ancilla would be driven in parallel with one of the bus stops.

To benchmark the shuttling performance, we use interleaved randomized benchmarking. The resonant control of qubit A1 when localized below P2 is used as the reference set. The interleaved operation consists of shuttling from below P2 to a location in the shuttling bus adjacent to a bus stop, idling for \SI{50}{\nano\second}, and shuttling back. A calibrated phase correction is added to make the targeted interleaved operation an identity gate. The measured fidelities do not change substantially with a small increase in the idling time, therefore we believe most of the infidelity originates during shuttling itself. A significant fraction of the shuttling error originates during loading and unloading the conveyor, when A1 moves from below gate P2 to below gate C0, as seen from the 97.9\% round-trip shuttling fidelity to bus stop 1. This transition is induced with a linear voltage ramp of \SI{40}{\nano\second} duration and may be possible to optimize further. Only a 0.2\% decrease in fidelity is observed when shuttling the remaining round-trip distance to bus stop 4. One round-trip of the ancilla from below P2 to bus stop 4, in the absence of any two-qubit interactions, takes place in \SI{730}{\nano\second}. Considering the $T_2^\mathrm{H}$ of A1 is \SI{47.6}{\micro\second} with a decay exponent of 1.7 when localized below P2, we estimate that dephasing alone could contribute about 0.1\% to the infidelity. Increasing the speed of the traveling-wave potential would be the most effective way to increase this limit (in~\cite{De_Smet_2025}, conveyor speeds were up to 25 times faster).

We did not observe any clear evidence of valley excitations limiting the quality of the multi-qubit demonstration. For example, such excitations could result in observing two closely-spaced resonance frequencies as spins occupying different valley-orbit states will have slightly different $g$-factors \cite{Kawakami_2014,Ferdous_2018}. This effect was observed during the initial tuning of the exchange coupling of bus stop 1, but subsequent changes to the loading pulse sequence removed the effect. 

It is understood that fluctuations in the valley splitting across a silicon shuttling channel can limit the quality of spin shuttling \cite{Losert_2024,Langrock2023,volmer2025reductionimpactlocalvalley}. It is possible that we would be able to resolve such effects by further optimizing the shuttling channel and rigorously characterizing the valley splitting across the array.

\section{Two-qubit characterization}
\label{supp:2Qcharacterization}

\begin{table}[]
    \centering
    \begin{tabular}{|c|c|c|c|c|}
        \hline
        Bus stop & Qubits & $F_\mathrm{ref}$ (\%) & $F_\mathrm{DCZ}$ (\%) & $F_\mathrm{Dsh}$ (\%) \\
        \hline
        1 & A1 and D1 & 96.2(2) & 84.2(8) & 91.9(5) \\
        2 & A1 and D2 & 97.4(3) & 87.9(6) & 93.0(4) \\
        3 & A1 and D3 & 96.7(2) & 82.8(1.3) & 92.6(3) \\
        4 & A1 and D4 & 97.7(1) & 89.3(5) & 93.9(8) \\
        \hline
    \end{tabular}
    \caption{Table of extracted fidelities using interleaved CRB. The reference fidelity $F_\mathrm{ref}$ corresponds to the fidelity of simultaneous single-qubit Clifford gates in the relevant two-qubit Hilbert space. The DCZ operation with fidelity $F_\mathrm{DCZ}$ is equivalent to the two-qubit version of the circuit displayed in Fig.~\ref{fig4}a. The decoupled shuttling operation with fidelity $F_\mathrm{Dsh}$ is the same two-qubit operation but with the entangling $CS$ interactions replaced by idling instead.}
    \label{tab:2QCRB}
\end{table}

We use character randomized benchmarking (CRB) to evaluate the performance of the two-qubit interaction between the shuttled ancilla and the four data qubits \cite{Xue_2019}. CRB allows us to use simultaneous single-qubit Clifford gates as the reference gate set as opposed to general two-qubit Clifford operations, which require many native CZ gates to compile.

As seen in Fig.~\ref{fig4}a, the entangling interaction used to implement parity checks is a composite operation consisting of coherent shuttling, two-qubit interactions, and single-qubit gates for refocusing phase pick-up. Furthermore, we take advantage of the fact that the native controlled-phase interactions commute. We therefore benchmark each individual DCZ interaction to estimate the isolated error contribution from each component of this composite interaction.

We perform two different rounds of interleaved CRB for each two-qubit interaction. First, only the decoupled shuttling sequence is interleaved with no exchange activated. As this operation should be logically equivalent to $X_{180}\otimes X_{180}$, this provides an estimate of the error rate associated with qubit shuttling and the refocusing gates in the two-qubit Hilbert space of the ancilla and the relevant data qubit. Then, we interleave a maximally-entangling DCZ interaction such that the interleaved gate is $X_{180}\otimes X_{180}\mathrm{diag}\left(1,1,1,-1\right)$. The inverting Clifford gate is implemented using a directly calibrated $CZ$ gate consisting of a single two-qubit interaction and explicitly calibrated phase corrections on both participating qubits. This allows us to use a verified lookup table to implement the inverting Clifford. We also verify with interleaved CRB that this direct CZ has comparable fidelity to the DCZ, in the range of 90\%-95\%. The decoupled version of the gate is conceptually closer to the compound interaction presented in Fig.~\ref{fig4}a used to implement parity checks in this work. Table~\ref{tab:2QCRB} summarizes all of the fidelities extracted from CRB.

Based on the interleaved CRB results, we coarsely estimate an error rate $r_\mathrm{Dsh} = 1-F_\mathrm{Dsh}$ associated with the decoupled shuttling sequence, and an error rate $r_\mathrm{DCZ} = 1-F_\mathrm{DCZ}$ associated with the full entangling two-qubit gate, including the evolution under exchange, shuttling and the refocusing pulses. By assuming these errors are stochastic, independent, and small, we approximate the error rate associated with the pure exchange component of the interaction as $r_J = r_\mathrm{DCZ}-r_\mathrm{Dsh}$. The fidelity of the two-qubit exchange interaction as reported in Fig.~\ref{fig3}g is then given as $F_\mathrm{2Q} = 1-r_J$.

\begin{figure*}
    \centering
    \includegraphics[width=\textwidth]{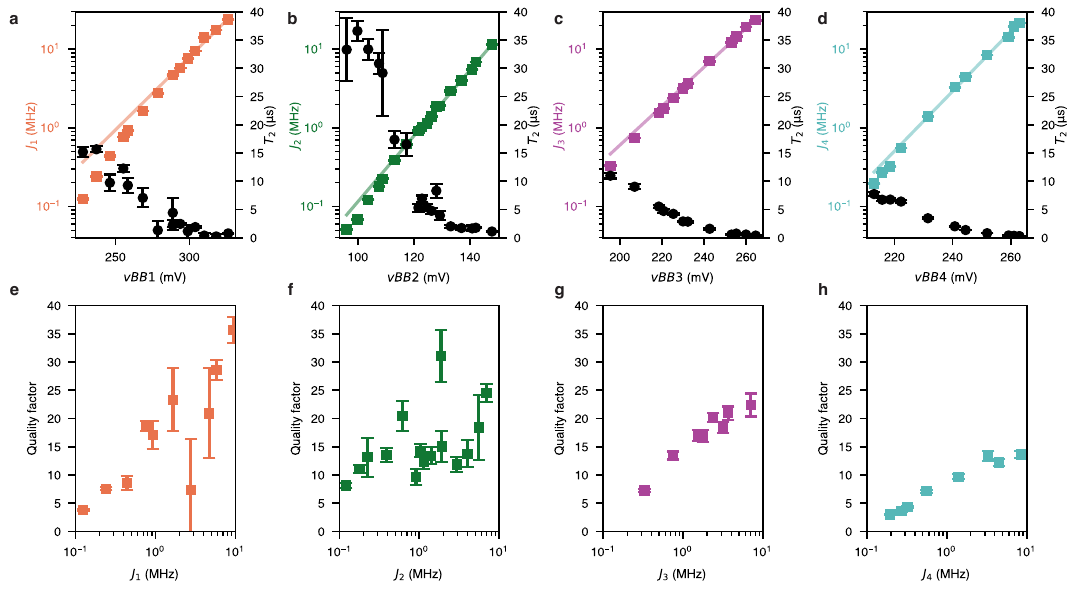}
    \caption{\textbf{a-d} Exchange strength and dephasing time of A1 as a function of barrier voltage for two-qubit interactions at the four bus stops respectively. \textbf{e-h} Quality factors for the two-qubit interactions at the four bus stops, calculated as $Q = 2JT_2$, estimating the number of maximally entangling two-qubit interactions that can take place within the characteristic decay time.}
    \label{fig:exchangeQ}
\end{figure*}

To verify these fidelity estimates, we can compare them to the quality factors obtained from observing exchange oscillations in a decoupled sequence. Fig.~\ref{fig:exchangeQ} summarizes the effect of increasing exchange on qubit coherence, and the extracted quality factors $Q$ for $1~\mathrm{MHz} < J < 10~\mathrm{MHz}$ fall in the range of 10 to 30. The quality factor puts a bound on the achievable fidelity limited by incoherent noise as $F_J < 1-1/Q$, since quasi-static noise is removed by the refocusing pulses when exchange oscillations are measured. This limit is consistent with the benchmarked fidelities of 90\%-95\% that are obtained from interleaved CRB.

\begin{table}[]
    \centering
    \begin{tabular}{|c|c|c|}
        \hline
        Operation & Duration & Error estimate \\
        \hline
        A1 $X_{90}$ & \SI{250}{\nano\second} & 0.0033 \\ 
        D1 $X_{90}$ & \SI{250}{\nano\second} & 0.0002 \\
        D2 $X_{90}$ & \SI{250}{\nano\second} & 0.0003 \\
        D3 $X_{90}$ & \SI{250}{\nano\second} & 0.0004 \\
        D4 $X_{90}$ & \SI{250}{\nano\second} & 0.0002 \\
        A1 and D1 exchange & \SI{440}{\nano\second} & 0.076 \\
        A1 and D2 exchange & \SI{220}{\nano\second} & 0.051 \\
        A1 and D3 exchange & \SI{360}{\nano\second} & 0.099 \\
        A1 and D4 exchange & \SI{360}{\nano\second} & 0.046 \\
        A1 round-trip shuttle & \SI{730}{\nano\second} & 0.023 \\
        A1 idling (excl. shuttling) & \SI{1.5}{\micro\second} & 0.003 \\
        D1 idling & \SI{3.65}{\micro\second} & 0.010 \\
        D2 idling & \SI{3.87}{\micro\second} & 0.020 \\
        D3 idling & \SI{3.48}{\micro\second} & 0.016 \\
        D4 idling & \SI{3.48}{\micro\second} & 0.010 \\
        A1 measurement & \SI{10}{\micro\second} & 0.017 \\
        \hline
    \end{tabular}
    \caption{Table summarizing the operations required to initialize a surface code logical state and their associated timescales and estimated error rates. For operations with a fidelity $F$ benchmarked with randomized benchmarking, the error rate is taken as $1-F$. The idling times $t_\mathrm{idle}$ are determined based on the nominal pulse sequence used to run the circuit shown in Fig.~\ref{fig4}c where single-qubit gates on D1/D2 and D3/D4 are parallelized. The idling errors are estimated using $1-\exp(-(t_\mathrm{idle}/T_2^\mathrm{H})^{\alpha'})$ where $T_2^\mathrm{H}$ are the measured Hahn echo times for each qubit (\SI{47.6(1.0)}{\micro\second}, \SI{43.2(1.0)}{\micro\second}, \SI{32.6(8)}{\micro\second}, \SI{37.8(7)}{\micro\second}, \SI{65.1(1.6)}{\micro\second} for A1, D1, D2, D3, D4 respectively) and $\alpha'$ is the measured echo decay exponent (1.71(9), 1.86(11), 1.83(13), 1.74(8), 1.58(10) for A1, D1, D2, D3, D4 respectively). The A1 measurement error is commensurate with a qubit visibility of 97\%.}
    \label{tab:errorbudget}
\end{table}

\begin{figure}
    \centering
    \includegraphics[width=\columnwidth]{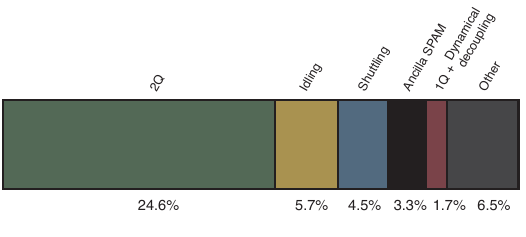}
    \caption{Graphical representation of the error budget for initializing a \(\left[\!\left[ 4, 1, 2 \right]\!\right]\) surface code logical state. The relative proportions are derived from the error estimates in Table~\ref{tab:errorbudget}.}
    \label{fig:errorbudget}
\end{figure}

\section{Error budget for logical state preparation}
\label{supp:errorbudget}

The combination of qubit characteristics (Fig.~\ref{fig3}e), operation benchmarks (Fig.~\ref{fig3}g), and circuit compilation can be used to construct an error budget for the 5-qubit processor. We choose the relevant example of preparing the logical $\ket{0_L} = \frac{1}{\sqrt{2}}\left(\ket{0000}+\ket{1111}\right)$ state for a \(\left[\!\left[ 4, 1, 2 \right]\!\right]\) surface code as it utilizes all operations. As shown in Fig.~\ref{fig4}f and Fig.~\ref{fig:rhos_Q4}, the sparse processor produces such a state with a fidelity of about 63\%. The quantum circuit producing this state, shown in Fig.~\ref{fig4}c, utilizes all four two-qubit interactions between the ancilla and each data qubit, two round-trips of shuttling the ancilla qubit, 16 decomposed single-qubit gates, and initialization and measurement of the ancilla qubit. The final single-qubit gates on each data qubit are compiled into the QST projections, and measurement errors associated with the data qubits are not considered as they are corrected during state reconstruction (see Sec.~\ref{supp:QST}). Table~\ref{tab:errorbudget} summarizes the error sources from all operations and idling taking place in the circuit.

Fig.~\ref{fig:errorbudget} represents the relative proportions of errors originating from single-qubit gates and dynamical decoupling, two-qubit interactions, shuttling, idling, and measurement. For a first-order estimate, we assume all errors $p_i$ are depolarizing such that the circuit produces a state $\rho=(1-P)\ket{\psi_\mathrm{GHZ}}\bra{\psi_\mathrm{GHZ}}+P\frac{I}{16}$ where $(1-P) = \Pi_i(1-p_i)$, yielding a logical state fidelity of about 67\%. A remaining error of about 6\% between the estimated and measured state fidelities is unaccounted for, but this can reasonably be expected to originate from sources that are not captured by the individual benchmarking protocols. These sources include crosstalk beyond the benchmarked Hilbert spaces and slow device drift. We believe the latter is particularly relevant for longer experiments. For example, collection of all tomographic projections of the five-qubit GHZ state shown in Fig.~\ref{fig:rhos_Q5} took 2 hours, during which all control parameters are held constant, and the measured state fidelity of 53\% is substantially lower than the four-qubit logical state initialization despite the circuits being very similar. Here, the solution is to interleave fine calibrations more densely through such experiments.

Although the presented analysis is coarse, the overall performance of the device is well-captured by the individual benchmarks. Crucially, we believe the most performance-limiting error sources are apparent. The two-qubit interaction fidelities represent the majority of the error present in the device, and the incoherent noise limiting them must be addressed. Fortunately, previous demonstrations of two-qubit gate fidelities well above 99\% in similar device architectures provide evidence that this is possible. The next largest error sources are idling and shuttling. These processes are dominated by dephasing (see Sec.~\ref{supp:1Qcharacterization}), meaning that increasing the speed of shuttling should suppress both sources of error.

\begin{figure*}
    \centering
    \includegraphics[width=\textwidth]{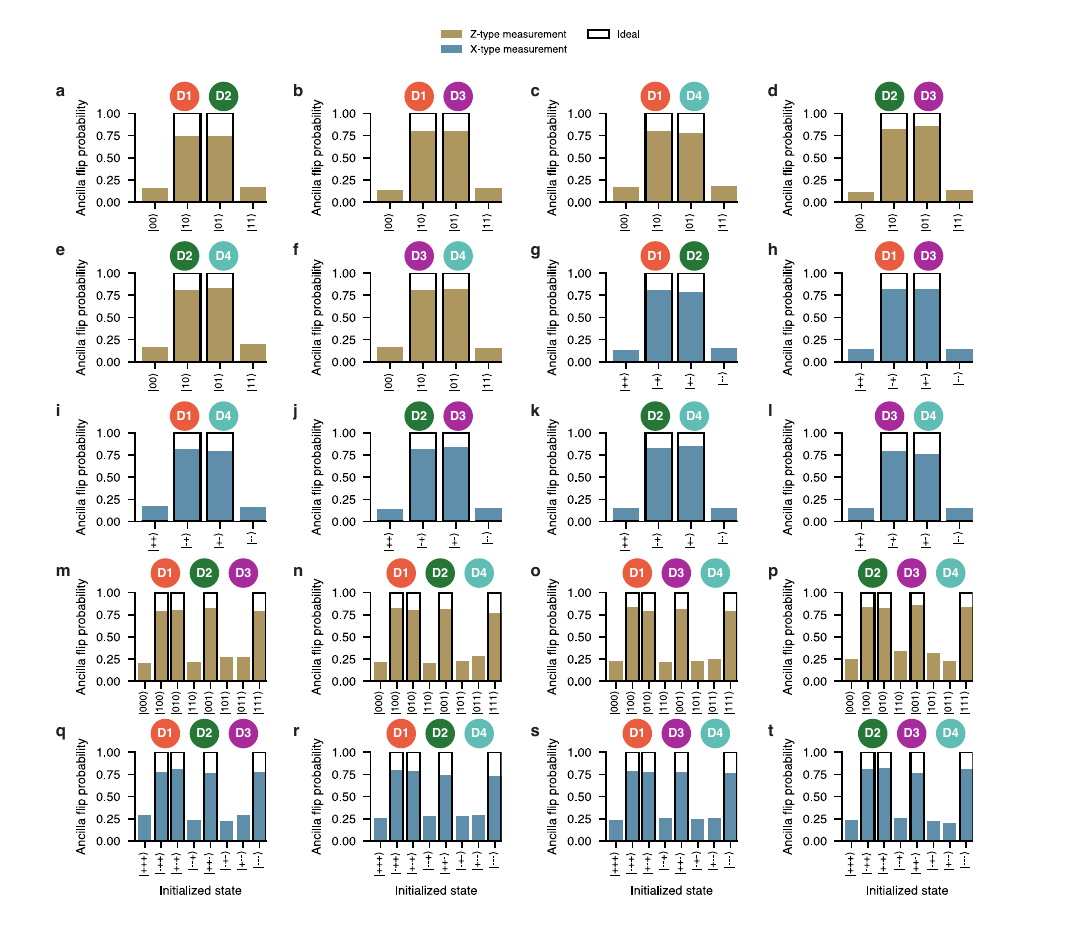}
    \caption{\textbf{a-f} Weight-two $Z$-type parity checks for all combinations of data qubits. \textbf{g-l} Weight-two $X$-type parity checks for all combinations of data qubits. \textbf{m-n} Weight-three $Z$-type parity checks for all combinations of data qubits. \textbf{q-t} Weight-four $X$-type parity checks for all combinations of data qubits.}
    \label{fig:paritychecks}
\end{figure*}

\begin{figure*}
    \centering
    \includegraphics[width=\textwidth]{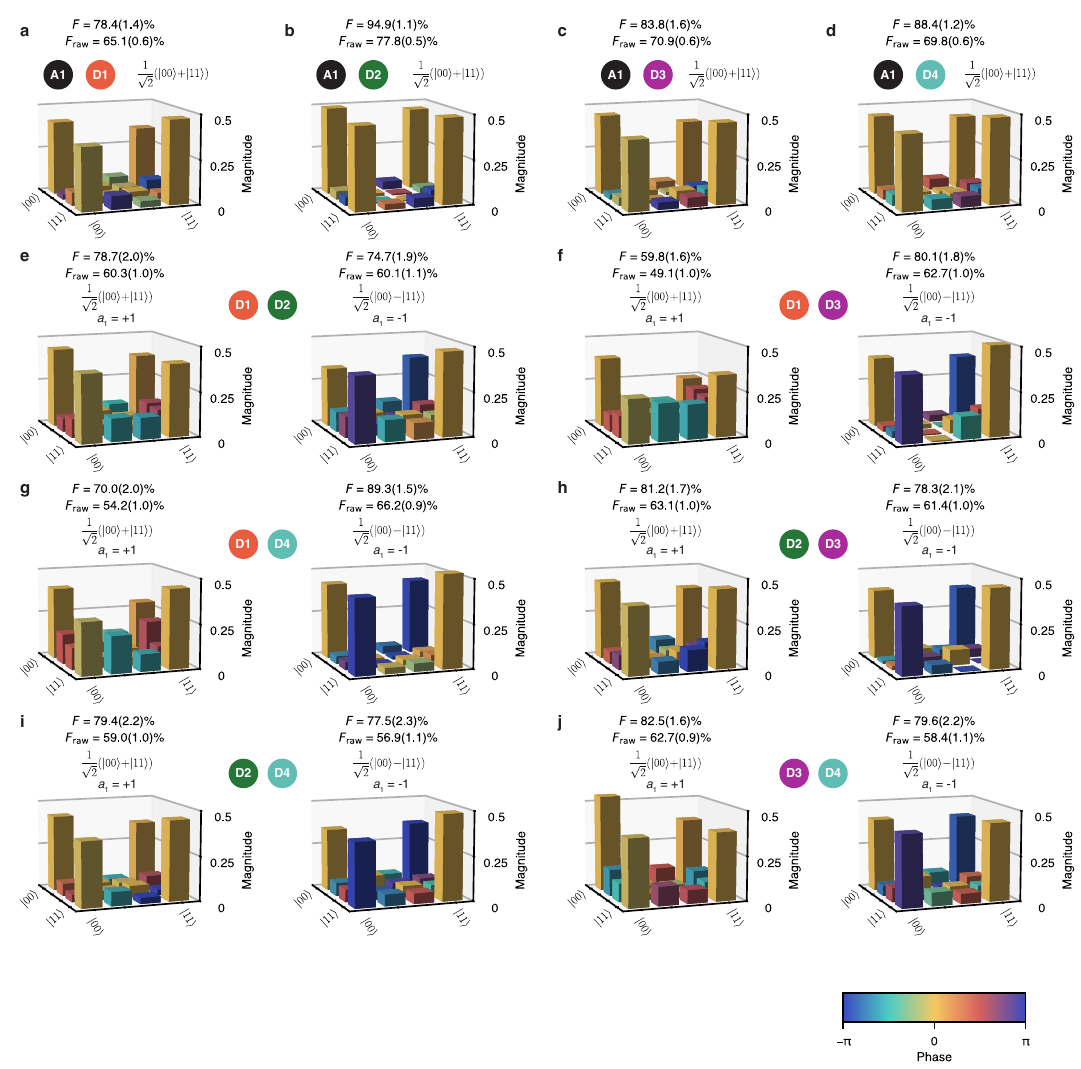}
    \caption{\textbf{a-d} Two-qubit GHZ state density matrices for all qubit combinations including the ancilla qubit A1. The states are prepared by running a circuit analogous to Fig.~\ref{fig4}c and performing QST on all qubits at the dashed line. \textbf{e-j} Two-qubit GHZ state density matrices for the displayed data qubit combinations. For each combination, two GHZ states differing in phase are probabilistically generated depending on the outcome of the weight-two $X$-type parity check used to create them as displayed above the plots. The fidelity $F = \bra{\psi_\mathrm{GHZ}}\rho_\mathrm{meas}\ket{\psi_\mathrm{GHZ}}$ is calculated with respect to the target state displayed next to each plot where $\rho_\mathrm{meas}$ is the reconstructed density matrix including corrections for readout errors. $F_\mathrm{raw} = \bra{\psi_\mathrm{GHZ}}\rho_\mathrm{raw}\ket{\psi_\mathrm{GHZ}}$ provides the analogous fidelity where $\rho_\mathrm{raw}$ is the reconstructed density matrix in the absence of readout corrections. Error intervals represent $\pm 1\sigma$ and are calculated by bootstrapping in all cases. Uncertainties in the readout corrections are included in the boostrapping procedure and therefore yield larger uncertainties for $F$ than $F_\mathrm{raw}$.}
    \label{fig:rhos_Q2}
\end{figure*}

\begin{figure*}
    \centering
    \includegraphics[width=\textwidth]{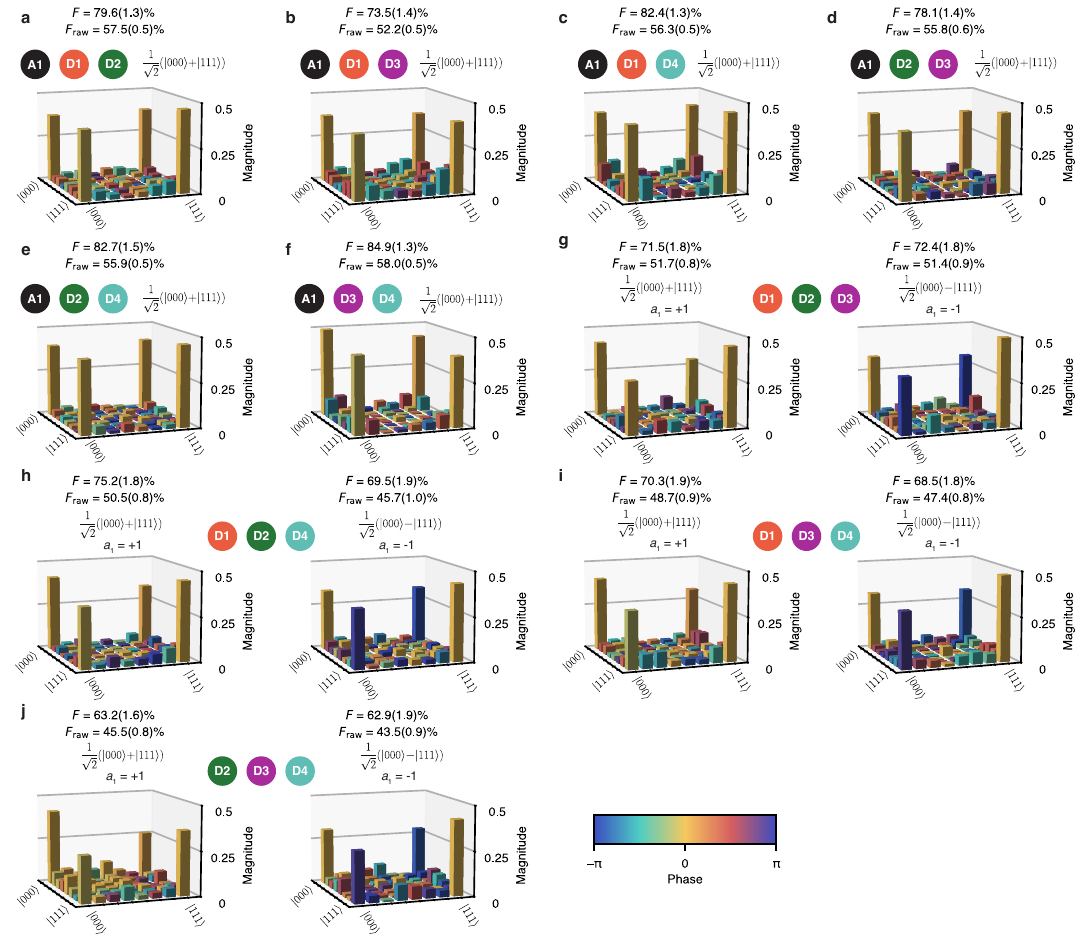}
    \caption{\textbf{a-f} Three-qubit GHZ state density matrices for all qubit combinations including the ancilla qubit A1. \textbf{g-j} Three-qubit GHZ state density matrices for the displayed data qubit combinations. For each combination, two GHZ states differing in phase are probabilistically generated depending on the outcome of the weight-three $X$-type parity check used to create them as displayed above the plots. The displayed fidelities are calculated as described by the caption of Fig.~\ref{fig:rhos_Q2}.}
    \label{fig:rhos_Q3}
\end{figure*}

\begin{figure*}
    \centering
    \includegraphics[width=\textwidth]{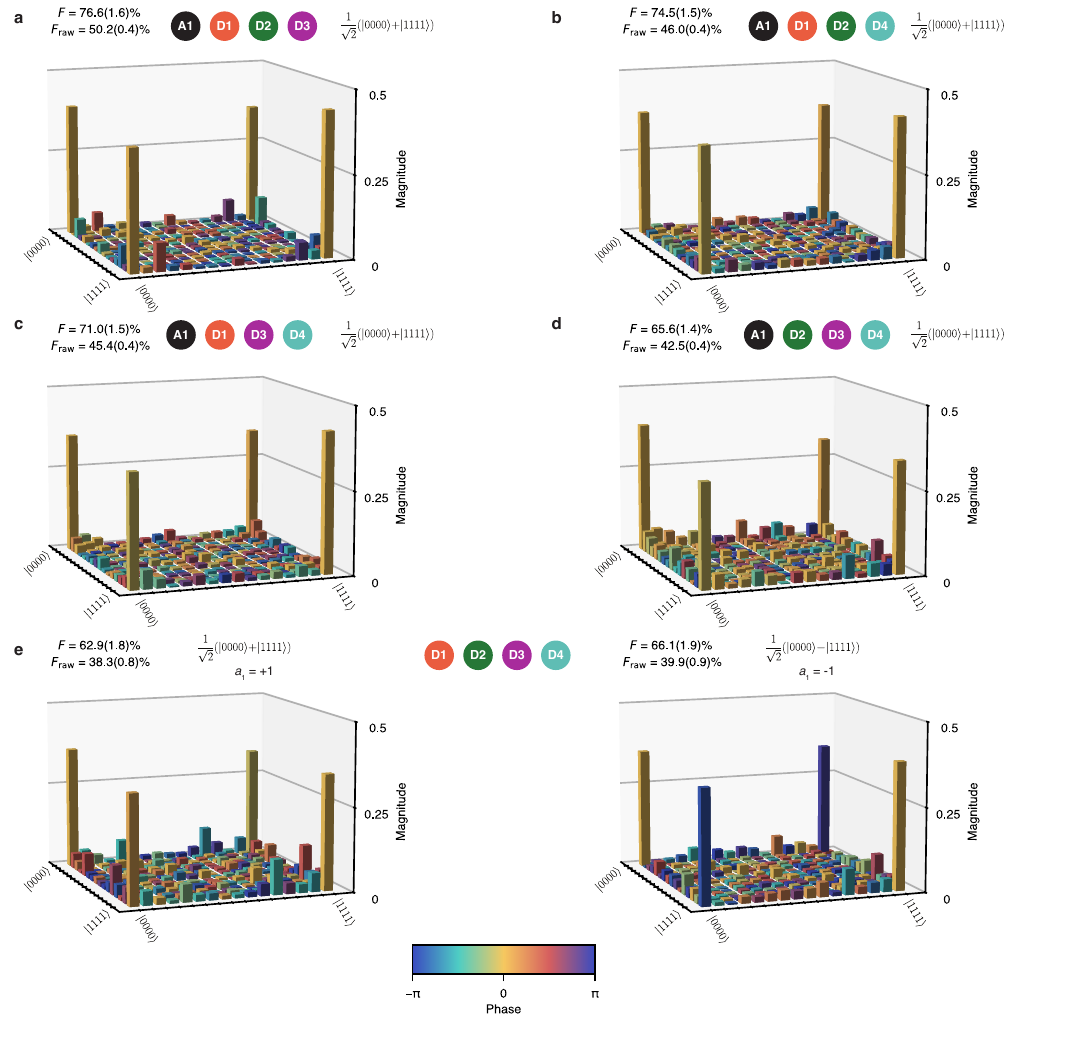}
    \caption{\textbf{a-f} Four-qubit GHZ state density matrices for all qubit combinations including the ancilla qubit A1. \textbf{g-j} Four-qubit GHZ state density matrices for all four data qubits. The two GHZ states differ in phase and are probabilistically generated depending on the outcome of the weight-four $X$-type parity check used to create them as displayed above the plots. The displayed fidelities are calculated as described by the caption of Fig.~\ref{fig:rhos_Q2}.}
    \label{fig:rhos_Q4}
\end{figure*}

\begin{figure*}
    \centering
    \includegraphics[width=\textwidth]{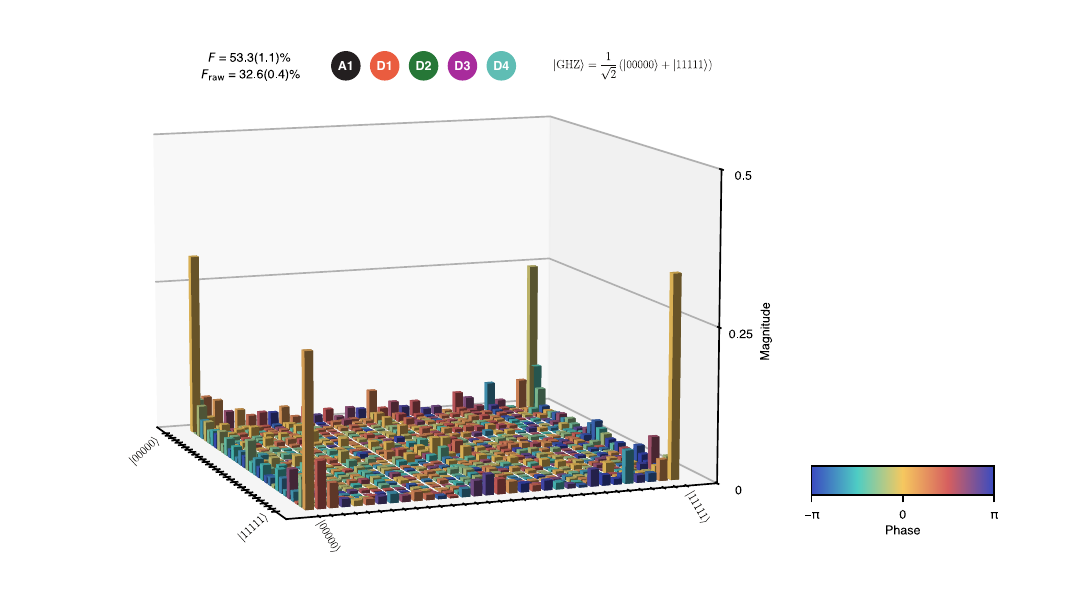}
    \caption{The five-qubit GHZ state density matrix supported by the ancilla qubit and all four data qubits. The displayed fidelities are calculated as described by the caption of Fig.~\ref{fig:rhos_Q2}.}
    \label{fig:rhos_Q5}
\end{figure*}

\section{Quantum state tomography}
\label{supp:QST}

QST for an $n$-qubit state is performed by measuring the expectation value of all $4^n$ Pauli observables $A_k$ using the individual computational-basis readout available for all qubits in the system. We bin the results into a probability vector $P_\mathrm{meas}$ of length $2^n$.

As the data qubits are read out indirectly and have finite visibility, readout errors are corrected by transforming $P_\mathrm{meas}$ according to the visibility of Rabi oscillations extracted using the same initialization and readout sequence \cite{Philips_2022}. We use a single round of QND measurement such that the data qubits have readout visibilities of about 80\% to 85\% (see Sec.~\ref{supp:QNDmeasurement}). If qubit $i$ has visibility limits $[V_{i,\mathrm{min}},V_{i,\mathrm{max}}]$, the corrected probability vector $P_\mathrm{corr}$ is given by:

\begin{equation}
    P_\mathrm{corr} = S_i^{-1}P_\mathrm{meas} = \begin{pmatrix}
        1-V_{i,\mathrm{min}} & 1-V_{i,\mathrm{max}} \\
        V_{i,\mathrm{min}} & V_{i,\mathrm{max}}
    \end{pmatrix}^{-1}P_\mathrm{meas}
\end{equation}

For multi-qubit corrections, the tensor product of the correction matrices $S_i$ for all participating qubits is used. The pseudo-inverse is used to calculate the inverse numerically to improve stability, the elements of $P_\mathrm{corr}$ are clipped to 1 and 0, and the vector is renormalized to maintain physicality (if applicable, these corrections are small and tend to lower the fidelity of the reconstructed state). The expectation value of the relevant operator is then calculated by weighting the entries of $P_\mathrm{corr}$ and averaging appropriately. For example, the expectation value $\langle XIY\rangle$ is extracted as $\mathrm{Tr}((Z\otimes I\otimes Z) \mathrm{diag}(P_\mathrm{corr}))$. When reconstructing states that include the ancilla qubit, we include its (relatively small) readout correction for consistency. For reconstructing states initialized via a parity check, no ancilla measurement correction is possible as data must be binned shot-by-shot. These reconstructed states therefore necessarily include errors in the ancilla measurement.

The density matrix $\rho_\mathrm{meas}$ is reconstructed from all measured expectation values $M_k$ using maximum likelihood estimation (MLE) to minimize $\sum_{k=1}^{4^n}\left|M_k-\mathrm{Tr}(A_k\rho_\mathrm{meas})\right|^2$ while enforcing the Hermiticity and unit-trace properties of $\rho_\mathrm{meas}$. The optimization is implemented using the CVXPY convex optimization package.

We use bootstrapping to estimate a statistical error on the extracted fidelity by performing the reconstruction multiple times by sampling $P_\mathrm{meas}$ from a multinomial distribution according to the measured probability and number of shots. We also uniformly sample the visibility limits used for readout correction, using the Rabi oscillation fit errors as bounds. Each state reconstruction is performed 500 times to ensure good convergence, and the mean fidelity is reported along with the $\pm 1\sigma$ standard deviation of the distribution of reconstructed state fidelities.

\begin{figure}
    \centering
    \includegraphics[width=\columnwidth]{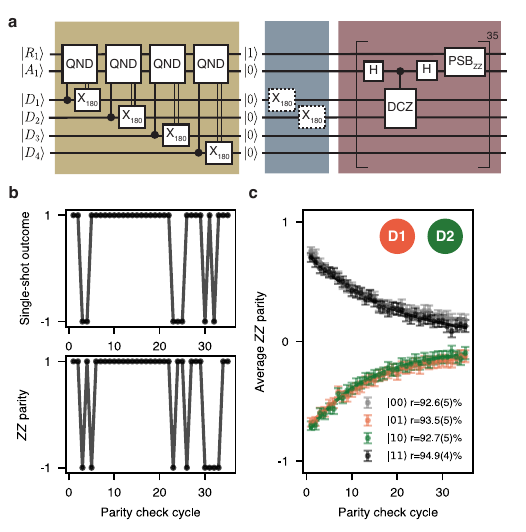}
    \caption{\textbf{a} Circuit for repetitive parity checks on bus stops 1 and 2. 
    \textbf{b} Trace of the single-shot outcomes and the corresponding parity values for a single run of the circuit with the $|00\rangle$ input state for bus stops 1 and 2. 
    \textbf{c} Average parity values for all input states $\ket{D2D1}$ as a function of the number of parity check cycles. The decay is fitted with $A r^{N} + C$, where $N$ is the number of rounds and $r$ is the parity retention rate. Error bar represents the 95\% confidence interval.}
    \label{fig:parity_check_analysis}
\end{figure}

\section{Parity check analysis}
\label{supp:paritychecks}

To estimate the accuracy of the parity checks as applied to eigenstates of the stabilizer, we use a similar process as for full state reconstruction with QST. After performing the weight-$w$ check and measuring the ancilla qubit, all data qubits are measured in the eigenbasis of the stabilizer. This yields a probability vector $P_\mathrm{meas}$ of length $2^{w+1}$. Readout errors on the data qubits are corrected as in Sec.~\ref{supp:QST} to acquire a corrected vector $P_\mathrm{corr}$. The ancilla qubit visibility is not corrected. We report the element of $P_\mathrm{corr}$ that corresponds to the intended data qubit state preparation and the correct ancilla measurement outcome. This yields a lower estimate for the parity check accuracy (Fig.~\ref{fig4}e) than an average of the correct ancilla outcome probabilities for all state preparations (Fig.~\ref{fig4}d), as it also accounts for instances where the parity check operation corrupted the data qubit state, even if the ancilla measurement gives the correct outcome.

We further assess the performance of a repetitive parity check circuit using bus stops 1 and 2. We initialize these bus stops in one of the states $|00\rangle, |01\rangle, |10\rangle$, or $|11\rangle$ and perform 35 rounds of $ZZ$ parity checks (Fig.~\ref{fig:parity_check_analysis}a). The measured parity is $+1$($-1$) if the first single-shot result returns a value of $+1$($-1$). Subsequent shots are compared with the previous shot to detect a flip or no flip in the parity; a trace of the single-shot values and the corresponding parity is shown in Fig.~\ref{fig:parity_check_analysis}b. We took 2,000 single-shots to obtain the average parity value for each parity check cycle. The decay of the average parity is fitted with the model $A r^{N} + C$, where $N$ is the number of rounds, to determine the parity retention rate $r$. The resulting retention rates are $r_{00}=92.64(5)\%$, $r_{01}=93.52(5)\%$, $r_{10}=92.70(5)\%$, and $r_{11}=94.89(4)\%$. 

\begin{figure*}
    \centering
    \includegraphics[width=\textwidth]{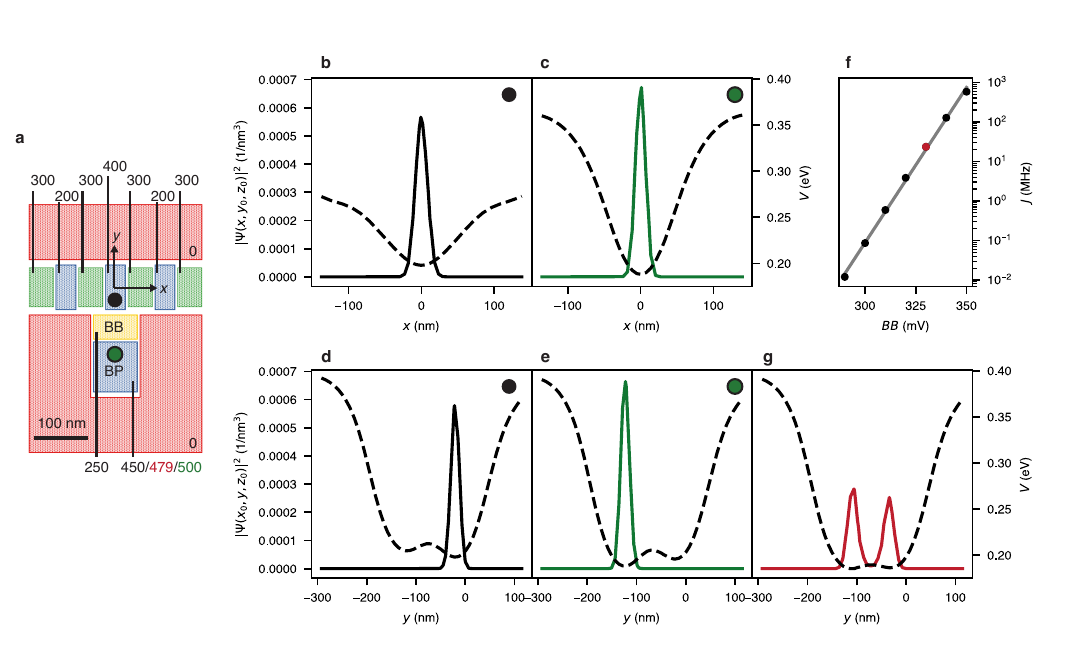}
    \caption{\textbf{a} Illustration of the boundary conditions used for electrostatic simulations of a double-dot potential at one of the bus stops. All numeric labels indicate a Dirichlet boundary condition in millivolts for the Poisson solver, which are set to approximate the potential landscape when a travelling-wave potential minimum is adjacent to the bus stop. Depending on the voltage applied to the plunger BP, the potential minimum is localized in the shuttling bus (black) or bus stop (green). \textbf{b} $x$-axis linecut of the potential and spatial wavefunction (probability density) of the ground state when the potential minimum is in the shuttling bus. \textbf{c} $x$-axis linecut of the potential and spatial wavefunction of the ground state when the potential minimum is in the bus stop. \textbf{d} $y$-axis linecut of the potential and spatial wavefunction of the ground state when the potential minimum is in the shuttling bus. \textbf{e} $y$-axis linecut of the potential and spatial wavefunction of the ground state when the potential minimum is in the bus stop. \textbf{f} Estimated relationship between barrier voltage $BB$ and exchange coupling as calculated from the perturbative relationship $J=4t_c^2/U$ where the onsite Coulomb potential $U\approx$~\SI{12}{\milli\eV} for all voltage settings. The linear fit corresponds to $J=J_\mathrm{off}\exp(0.18BB/\mathrm{mV})$. \textbf{g} $y$-axis linecut of the potential and spatial wavefunction of the ground state when the double-dot potential is balanced at the voltage setting highlighted in \textbf{f}. The orbital energy spacing to the first excited state is equal to $2t_c$.}
    \label{fig:electrostatics}
\end{figure*}

\section{Electrostatic design}
\label{supp:electrostatics}

We use QTCAD to perform Poisson and Schr\"{o}dinger simulations on a simplified device model \cite{Beaudoin_2022}. We treat these simulations as a qualitative confirmation of intuition rather than an attempt at rigorous device modeling. We use a subset of the nominal gate design for a bus stop as shown in Fig.~\ref{fig:electrostatics}a and perform a 3D Poisson simulation of the electrostatics using the indicated gate voltages as Dirichlet boundary conditions. These voltages are physically reasonable but generally smaller than those used in experiment due to screening effects which are not considered here. The bounding area for the simulation is kept large enough such that the natural boundary conditions do not affect the converged solution. The electrostatic solution is used as the static potential for a 3D time-independent Schr\"{o}dinger equation solver to find the eigenspectrum of a single-electron in the effective double-dot potential.

\begin{figure*}
    \centering
    \includegraphics[width=\textwidth]{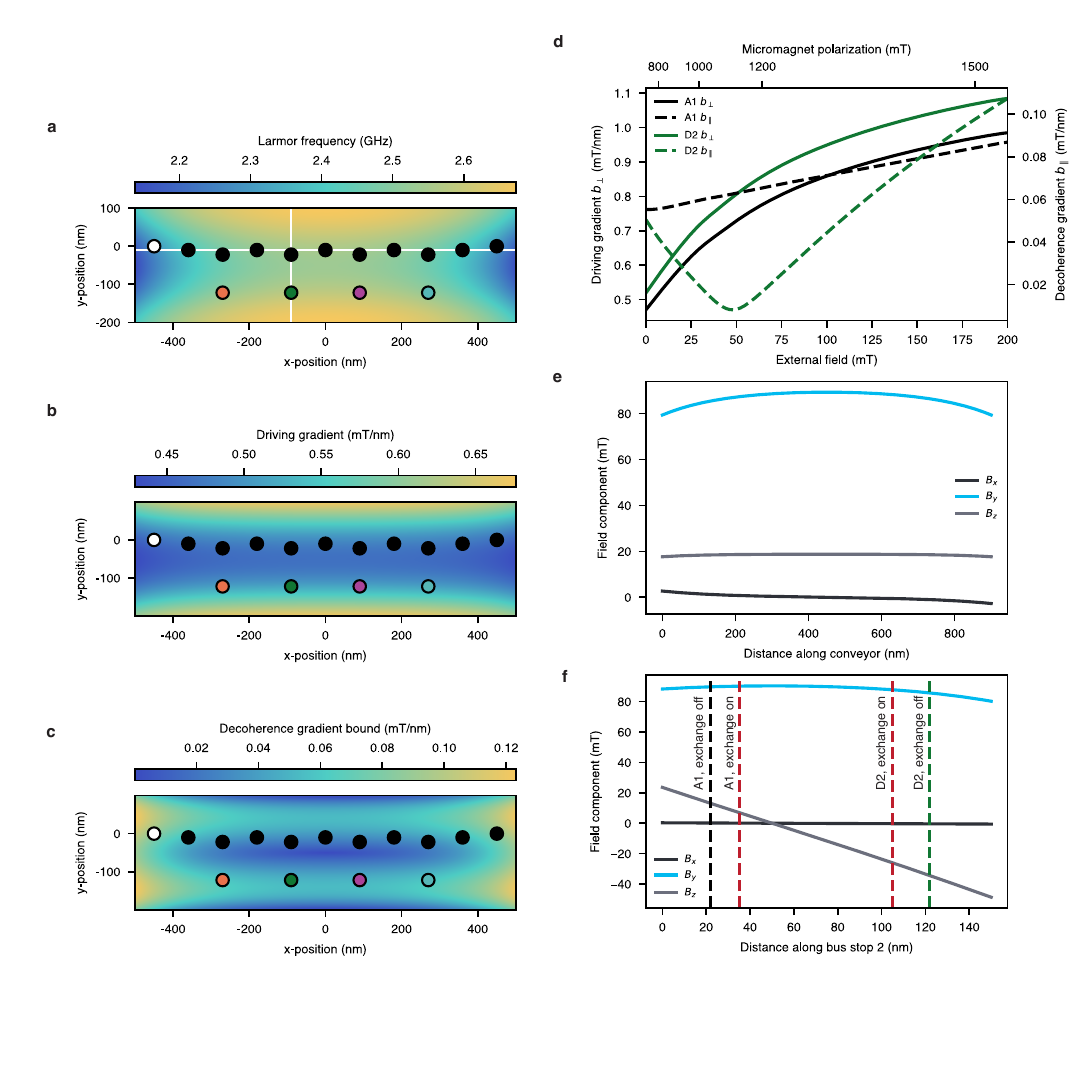}
    \caption{\textbf{a} Simulated Larmor frequencies of an electron spin (with $g=2$) in the quantum well when $B_\mathrm{ext}=0$ after partial polarization of the micromagnet. Points represent estimated locations of the spin localized along the shuttling bus and the bus stops as a guide to the eye. \textbf{b} Simulated driving gradient in the quantum well when $B_\mathrm{ext}=0$ after partial polarization of the micromagnet, assuming a control field oriented along the $y$-axis. \textbf{c} Simulated decoherence gradient in the quantum well when $B_\mathrm{ext}=0$ after partial polarization of the micromagnet. \textbf{d} The estimated impact of the external field setting, and therefore micromagnet polarization, on the driving and decoherence gradients at two representative quantum dot locations in the array. A1 is assumed to be localized below gate P2, as this is where it is resonantly controlled and idles in our experiments. Generally, the driving gradient decreases monotonically with external field, while the decoherence gradient may exhibit non-monotonic behavior. The homogeneous polarization of the micromagnet, in units \SI{}{\milli\tesla}, is phenomenologically fit to $1.155\times 10^{-4}x^3-5.515\times 10^{-2}x^2+10.38x+734.1~[\mathrm{mT}]$ where $x = B_\mathrm{ext}/\mathrm{mT}$ based on the measured Larmor frequencies of R1 and A1 as the external field is lowered from \SI{200}{\milli\tesla} to zero. \textbf{e} Simulated magnetic field components along the shuttling bus axis as indicated by the horizontal white line in \textbf{a} when $B_\mathrm{ext}=0$ after partial polarization of the micromagnet. The maximum estimated tip in magnetic field vector along this axis is \SI{2}{\degree}. \textbf{f}  Simulated magnetic field components along the double-dot potential of bus stop 2 as indicated by the vertical white line in \textbf{a}. The vertical dashed lines indicate estimated spin qubit locations depending on whether the exchange interaction is pulsed on or off. The estimated tip in magnetic field vector between conveyor and bus stop dots ranges between \SI{30}{\degree} when exchange is off and \SI{20}{\degree} when exchange is on.}
    \label{fig:magnetostatics}
\end{figure*}

Fig.~\ref{fig:electrostatics}b-e illustrates $x$- and $y$-axis linecuts of the ground state when a single-electron is confined in the effective double-dot potential formed when the traveling-wave potential minimum is positioned adjacent to the bus stop. The simulated potential minima shift orthogonally to the axis of the shuttling bus as a result of the presence of the bus stops, as illustrated in Fig.~\ref{fig:magnetostatics}a. Depending on the bus stop plunger voltage BP, the electron's orbital ground state is either localized in the shuttling bus or the bus stop. We can therefore extract estimated positions of the spin qubits as input to magnetostatic simulations as detailed in Sec.~\ref{supp:magnetostatics}. Furthermore, we extract orbital energy splittings of \SI{3.8}{\milli\eV} for the shuttled dot and \SI{4.1}{\milli\eV} for the bus stop dot which are relevant for EDSR. Using a simplified many-body simulation, the onsite Coulomb potential is estimated to be \SI{12}{\milli\eV} for both dots.

The microscopic valley degree of freedom is not accounted for in the simulations. Nevertheless, the tunability of the bare tunnel coupling $t_c$ may be extracted from the energy difference $2t_c$ between bonding and antibonding states when the double-dot potential is balanced. An example is shown in Fig.~\ref{fig:electrostatics}g. The bare exchange coupling follows from perturbation theory as $J=4t_c^2/U$~\cite{burkard_2023}. The true exchange coupling in silicon will be influenced by the valley phase difference between the occupied eigenstate of each dot. However, $J\propto t_c^2$ remains true. Therefore, the barrier lever arm can be estimated by fitting to a model $J=J_\mathrm{off}\exp(\alpha BB/\mathrm{mV})$ where $BB$ is the barrier gate voltage modulating exchange at a balanced double-dot potential. As shown in Fig.~\ref{fig:electrostatics}f, we extract $\alpha=0.18$ for exchange tunability in the model. This can be compared to the measured relations $\alpha_1 = 0.04$, $\alpha_2 = 0.10$, $\alpha_3=0.06$, and $\alpha_4=0.09$ for bus stops 1 to 4 respectively. The reasonable qualitative agreement provides a basis for improving design rules that will become increasingly relevant for larger device designs.

\section{Magnetostatic design}
\label{supp:magnetostatics}

The micromagnet is deposited with a window shape (see Fig.~\ref{fig1}a) such that its stray field in the quantum well gives rise to a spin Hamiltonian that is practical for measuring and interacting qubits in the sparse array while being reasonably robust to small fabrication defects and microscopic variation. The design criteria are to permit single-spin control with EDSR at all sites in the array with addressability while maintaining resilience against electric charge noise. EDSR controllability enters the Hamiltonian through $b_\perp$, the magnetic field gradient transverse to the total field vector at each potential quantum dot location, and charge noise susceptibility is quantified by $b_\parallel$, the longitudinal magnetic field gradient parallel to the total field vector. $b_\parallel$ also gives rise to addressability between spins and the Zeemen energy differences required for high-fidelity parity-mode PSB readout \cite{Seedhouse_2021} and fast adiabatic exchange interactions for two-qubit gates \cite{Rimbach-Russ_2023}.

In order to calculate the relevant quantities, the point-like locations of the relevant quantum dots are extracted from electrostatic simulations as described in Sec.~\ref{supp:electrostatics}, qualitatively approximating the dot positions in the experiment. We use the magpylib Python package to estimate the stray field of the micromagnet assuming homogeneous polarization, which is an approximation \cite{Aldeghi2025}. We use a magnet polarization of \SI{730}{\milli\tesla} to match the simulated qubit Larmor frequencies to the values measured in experiment. As the saturation polarization for cobalt micromagnets has been observed in the range of \SIrange{1.5}{1.8}{\tesla}, the magnet is partially demagnetized. However, we do not observe signs of micromagnet instability over several months of experiments. A detailed description of extracting the relevant quantities can be found in \cite{Unseld2025}.

Fig.~\ref{fig:magnetostatics}a shows the total magnetic field in the quantum well, expressed as the spin Larmor frequency $f_\mathrm{L}(x,y)$ assuming a $g$-factor of 2. The theoretical distribution shown in Fig.~\ref{fig1}c is taken along the dashed line, while the estimates for quantum dots localized in the bus stops are taken using the maximum of the simulated single-particle spatial wave functions. For three of the four bus stops, we measure the qualitatively-expected Zeeman energy difference between bus and bus stop spins. Bus stop 4 is an exception, where the bus stop spin is lower in frequency than the bus, and the Larmor frequencies only differ by about \SI{20}{\mega\hertz}. This requires a slower two-qubit interaction to maintain adiabaticity. The exact reason for this discrepancy presumably depends on the microscopics of the device which we cannot deduce with certainty. The micromagnet design is quite robust to misalignments with the external field (i.e. the behavior is nearly unchanged if the magnetization vector does not point exactly along the $y$-axis). Therefore, it may be a result of either magnetic domain alignment in the micromagnet, which our simulations do not account for; a misalignment in the deposition of the micromagnet; or a shift in the actual position of the spins with respect to their estimated locations on the order of tens of nanometers along the $y$-axis. In fact, we rely on some degree of disorder, as the nominal micromagnet design with homogeneous polarization gives rise to a perfectly symmetric profile such that data qubits 1 and 4 as well as 2 and 3 should have identical frequencies. It is anticipated from previous experiments that microscopic variation will lift this degeneracy on the order of tens of megahertz, which is what we observe in experiment.

The estimated transverse gradient $b_\perp(x,y)$ for driving along the $y$-axis is shown in Fig.~\ref{fig:magnetostatics}b and is dominated by the $\partial B_z(x,y)/\partial y$ gradient component. It is relatively constant over the sparse array, and we expect similar Rabi frequencies $f_\mathrm{R}$ to be achieved for equivalent driving amplitudes for all qubits. As $f_\mathrm{R}$ also depends on the microscopic dipole moment of the driven quantum dots (see Sec.~\ref{supp:universalcontrol}), the driving amplitudes required to achieve a constant $f_\mathrm{R}$ across the array typically vary by an order-unity factor.

The longitudinal gradient $b_\parallel(x,y)$ through which electric noise couples to the spin depends on the spatial orientation of the fluctuation. We therefore take $\left|\nabla f_\mathrm{L}(x,y)\right|$ as an estimate of the noise susceptibility, which will consist of primarily $\partial B_y(x,y)/\partial x$ and $\partial B_y(x,y)/\partial y$ components. Fig.~\ref{fig:magnetostatics}c shows the sweet spot that emerges along the $x$-axis of symmetry. While this sweet spot cannot be shared by both the bus axis and the bus stops, in all cases the gradient remains below \SI{0.1}{\milli\tesla/\nano\meter} which is considered reasonable \cite{Yoneda_2017, Philips_2022}.

Since the superconducting solenoid we used cannot be operated in persistent mode, the device is operated with only the remanence field of the micromagnet to avoid magnetic noise arising from current noise in the supply that drives the solenoid. The inclusion of a noiseless external field would have meaningful consequences. Two illustrative cases are plotted in Fig.~\ref{fig:magnetostatics}d for the ancilla qubit A1 localized below P2 and data qubit D2 localized below BP2. In general, a monotonic decrease in driving gradient is observed as the external field is lowered to zero. The decoherence gradient generally decreases as the external field is lowered from \SI{200}{\milli\tesla}, but the trend is not necessarily monotonic and the minimum may not be located at zero external field. In fact, a performance boost would likely follow from operating with a noiseless external field of \SI{50}{\milli\tesla} oriented along the $y$-axis (instead of zero external field) as both the Rabi frequency should increase and most qubits would be more robust to charge noise.

Fig.~\ref{fig:magnetostatics}e shows the magnetic field components along the shuttling bus and highlights the homogeneity of the estimated Larmor vector across the array. The orientation of the magnetic field vector at one end of the conveyor differs from that at the center by only about \SI{2}{\degree} which facilitates fast adiabatic spin shuttling. Fig.~\ref{fig:magnetostatics}f shows the magnetic field components along the axis of bus stop 2 that are relevant for the two-qubit interaction taking place there. A sizable tip of about \SI{25}{\degree} exists between the magnetic field vectors at each quantum dot location. Such a tip gives rise to both spin-conserving and spin-flipping tunnel coupling between quantum dots. Along with the moderate expected Zeeman energy difference of about \SI{50}{\mega\hertz} between the interacting qubits, this makes the adiabatic controlled-phase gate more appealing to implement than other possible two-qubit interactions. If the external field were made to oppose the micromagnet polarization, the spin Larmor frequencies would decrease and the tip angle between bus stop and shuttling bus dots would increase, and baseband hopping control at the bus stop locations would also be possible \cite{Wang_2024, Unseld2025}.

\bibliography{references_noDOI}

\end{document}